%% file: main.tex
\documentclass{article}
\usepackage[english]{babel}
\usepackage[super,comma,sort&compress]{natbib}
\usepackage{url}
\makeatletter
\renewcommand\@biblabel[1]{#1} 
\makeatother

\usepackage{authblk}
\usepackage{amsthm}
\usepackage{amsmath}
\usepackage{amsfonts}
\usepackage{natbib}
\usepackage{amssymb}

\theoremstyle{plain}

\usepackage{graphicx}
\usepackage{algorithm}
\usepackage{algpseudocode}
\usepackage{pgfplots}
\usepgfplotslibrary{dateplot}

\usepackage{xspace}

\newcommand{\indep}{\perp \!\!\! \perp}
\usepackage{xcolor}
\usepackage{bbding}

\usetikzlibrary{shapes,decorations,arrows,calc,arrows.meta,fit,positioning}
\tikzset{
    -Latex,auto,node distance =2 cm and 2 cm,semithick,
    state/.style ={ellipse, draw, minimum width = 0.7 cm},
    point/.style = {circle, draw, inner sep=0.04cm,fill,node contents={}},
    bidirected/.style={Latex-Latex,dashed},
    el/.style = {inner sep=2pt, align=left, sloped}
}
\usepackage{caption}
\usepackage{subcaption}

\usepackage{float}
\usepackage{stfloats}
\usepackage{lineno}
\usepackage{stfloats} 
\usepackage{enumitem,wrapfig}
\usepackage[colorlinks,citecolor=blue,urlcolor=blue,filecolor=blue,backref=page]{hyperref}

\begin{document}

\title{Estimate an Exposure Response Function with Negative Controls: A Bayesian Nonparametric Approach}

\author[* $\dagger$]{Jie Kate Hu}
\author[* $\ddagger$]{Dafne Zorzetto}
\author[$\dagger$]{Francesca Dominici}
\affil[$\dagger$]{Department of Biostatistics, Harvard T.H. Chan School of Public Health}
\affil[$\ddagger$]{Department of Statistics, University of Padova}
\affil[*]{These two authors contributed equally to this work, alphabetically ordered by surnames}


\maketitle

\begin{abstract}
Unmeasured confounding bias threatens the validity of observational studies.  While sensitivity analyses and study designs have been proposed to address this issue, they often overlook the growing availability of auxiliary data. Using negative controls from these data is a promising new approach to reduce unmeasured confounding bias.   In this article, we develop a Bayesian nonparametric method to estimate a causal exposure-response function (CERF) leveraging information from negative controls to adjust for unmeasured confounding. We model the CERF as a mixture of linear models. This strategy captures the potential nonlinear shape of CERFs while maintaining computational efficiency, and it leverages closed-form results that hold under the linear model assumption.  We assess the performance of our method through simulation studies. We found that the proposed method can recover the true shape of the CERF in the presence of unmeasured confounding under assumptions. To show the practical utility of our approach, we apply it to adjust for a possible unmeasured confounder when evaluating the relationship between long-term exposure to ambient $PM_{2.5}$ and cardiovascular hospitalization rates among the elderly in the continental US. We implement our estimation procedure in open-source software and have made the code publicly available to ensure reproducibility.
\end{abstract}

{\bf{Keywords}}: bias; confounding; negative controls; Bayesian analysis; nonparametrics

\section*{INTRODUCTION}
Negative controls (NC)  is a powerful tool to detect and adjust for unmeasured confounder bias \cite{tchetgen_2020_introduction, miao2018identifying} in observational studies. The key idea of NC analysis is to use auxiliary data that contain valuable information on unmeasured confounding mechanisms to estimate causal effects. Negative control exposure (NCE) and negative control outcome (NCO) variables are examples of such auxiliary information. An NCE variable $Z$ is a variable known not to cause an outcome $Y$ and an NCO variable $W$ is a variable known not to be causally affected by an exposure $X$ \cite {lipsitch2010negative}.  Under the assumption that
the association between the NCE variable $Z$ and $Y$ is subject to the same unmeasured confounding mechanisms as the association between exposure $X$ and $Y$,  the detection of an association between $Z$ and $Y$ can signal the presence of an unmeasured confounder $U$ (Fig. \ref{fig:NC}). Similarly, under the assumption that the association between $X$ and the NCO variable $W$ is subject to the same unmeasured confounding mechanism as between $X$ and the outcome $Y$, the detection of an association between $X$ and $W$ can signal the presence of an unmeasured confounder $U$ (Fig. \ref{fig:NC}). Sometimes an accurate measurement of a confounder is not available and the best we can often obtain is its proxies. These proxies are informative about the true unmeasured confounder and can be also treated as NCEs or NCOs for unmeasured confounding adjustment \cite{tchetgen_2020_introduction}.

\begin{figure}[!pb]
    \centering
    \begin{minipage}{0.9\linewidth}\centering
    \scalebox{0.9}{
\begin{tikzpicture}
    \node[state] (1) {$X$};
    \node[state] (2) [right =of 1] {$Y$};
    \node[state] (3) [above left =of 1, yshift = -1cm] {$U$};
    \node[state] (4) [above =of 1] {$Z$};
    \node[state] (5) [right =of 4] {$W$};
    \path (1) edge node[below] {} (2);
    \path (3) edge node[below] {} (1);
    \path (3) edge node[above] {} (2);
    \path (3) edge node[above] {} (4);
    \path (3) edge node[above] {} (5);
\end{tikzpicture}}
\end{minipage}
\hspace{1cm}
\begin{minipage}{0.9\linewidth}
\vspace{0.5cm}
\begin{itemize}
    \item W: negative control outcome
    \item Z: negative control exposure
    \item U: unmeasured confounder 
    \item X: treatment or exposure 
    \item Y: outcome
\end{itemize}
\vspace{1cm}
\end{minipage}
    \caption{An illustration of negative control exposure and outcome.}
    \label{fig:NC}
\end{figure}

Miao et al.\cite{miao2018identifying} and Cui et al.\cite{cui2023semiparametric} have demonstrated that, under certain conditions, the presence of a pair of NCE and NCO guarantees nonparametric identification of the average treatment effect (ATE) in the presence of unmeasured confounding. Many methods have been developed to estimate the ATE for binary and categorical treatments with NC \cite{cui2023semiparametric, shi2020multiply, ghassami2022minimax}. In this article, we focus on continuous treatments or exposures (such as drug dose, air pollution levels, and duration).

The causal effect of continuous exposures on a given outcome is often described by the causal exposure-response function (CERF) or dose-response curve. Several machine learning methods have been developed in the literature to estimate the CERF nonparametrically with NC under unmeasured confounding. Examples include two-stage kernel ridge regression \cite{singh2023kernel},  two-stage regression with adaptive basis derived from neural networks \cite{xu2021deep},  the maximum moment restriction method \cite{mastouri2023proximal},  neural maximum moment restriction method \cite{kompa2022deep}, and the minimax learning estimation method \cite{kallus2022causal}. Among these, Mastouri et al. \cite{mastouri2023proximal} and Kompa et al. \cite{kompa2022deep}  compared the performance of some of these techniques. While many of these methods have been proven to consistently estimate the CERF, most performed poorly in simulation studies with continuous treatments when evaluated on finite samples\cite[(see Supplementary Figures S3 - S6 in Kompa et al][]{kompa2022deep}). Moreover, these machine learning algorithms  rarely provide uncertainty bounds for the CERF estimates. 

In this article, we propose a new nonparametric approach to estimate the potentially nonlinear CERF of a continuous exposure ($X$) on a continuous outcome ($Y$) in the presence of unmeasured confounders ($U$). We make the following contributions: 1) we develop a new Bayesian nonparametric approach (BNP) that addresses unmeasured confounding using NC for the analysis of observational data; 2) we quantify the uncertainty of our estimates in the form of credible intervals (CI); and 3) we make the open-source software code publicly available.

\section*{A BRIEF REVIEW}
To estimate the CERF of a continuous exposure ($X$) on a continuous outcome ($Y$) in the presence of unmeasured confounders ($U$),  we develop a new BNP approach that addresses unmeasured confounding using NC. Before introducing our method, we review the conditions required to identify the causal effect of $X$ on $Y$ under linear models and the setup of BNP mixture models. Our method is built upon these two methods, but extends them to estimate the CERF nonparametrically. 

Throughout the paper,  we consider a sample of i.i.d. observations on exposure $X$, outcome $Y$, NCE variable $Z$, and NCO variable $W$. We denote the unmeasured confounder by $U$ and let $Y(x)$ be the potential outcome observed if $X$ were set to the value $x$.

\subsection*{Identification of average treatment effect}
Shi et. al\cite{shi_selective_2020} outlined the conditions for identifying the causal effect when the observed variables $(Y, X, Z, W)$ and the unobserved variable $U$ follow the linear models below:
\begin{equation}\label{eqn: shi}
\begin{array}{l@{}r}
    E[Y | X=x, U] = \beta_{Y0} + \beta_{YX}x + \beta_{YU}U, \\
    E[W | U] = \beta_{W0} + \beta_{WU}U, \\
    E[U | X=x, Z] = \beta_{U0} + \beta_{UX}x + \beta_{UZ}Z.
\end{array}
\end{equation}
More specifically, they considered the causal effect as a constant difference in the potential outcomes between two exposure levels and provided the following conditions to identify the causal effect under linear models:
\begin{itemize}         
     
     \item C1. (Z is an NCE): $Y(x, z) = Y(x)$ and  $Z\indep Y(x)|U,X$;
    \item C2. (W is an NCO): $W(x, z) = W$ and $W \indep X|U$;
    \item C3. (NCE and NCO are conditionally independent): $Z\indep W|U$;
\item C4 ($W$ and $Z$ are informative  about $U$): $\beta_{WU} \neq 0$ and $\beta_{UZ} \neq 0$.
\end{itemize}
Figure \ref{fig:NC} is one causal diagram among many that satisfy these assumptions. Consider fitting the following linear models to the observed data: 
\begin{eqnarray*}
E[Y | X, Z] &=& \theta_{Y0} + \theta_{YX}X + \theta_{YZ}Z \\
E[W | X, Z] &=& \theta_{W0} + \theta_{WX}X + \theta_{WZ}Z.
\end{eqnarray*}
Shi, et.al\cite{shi_selective_2020} showed when the NCE variable $Z$ and the NCO variable $W$ satisfied  conditions $C1-C4$ and equations in \eqref{eqn: shi}
\begin{equation}\label{eqn: beta}
    \beta_{YX} = \theta_{YX} - \theta_{WX}\frac{\theta_{YZ}}{\theta_{WZ}}
\end{equation}
in which $\theta_{WZ} = \theta_{WU}\theta_{UZ} \neq 0$ by condition C4. When $E[Y(x)] =  E[Y|X =x,  U)]$,  \eqref{eqn: shi}  implies $\beta_{YX}$ is their causal effect of interest
\begin{eqnarray*}
\beta_{YX} &=& E[Y(x+1)] - E[Y(x)] \\
&=& E[Y|X =x+1,  U)]-  E[Y|X =x,  U)] ;
\end{eqnarray*}
It is identifiable in the closed form of \eqref{eqn: beta}.

\subsection*{Probit stick breaking process}
\label{subsec:probit}
A general BNP mixture model assumes
 \begin{equation}\label{mod: Y}
Y | G\sim \int f(\cdot|\phi)G(d\phi)
 \end{equation}
where $f(\cdot|\phi)$ is a given parametric kernel indexed by parameter $\phi$, G is a mixing distribution,
which is assigned a 
flexible prior \cite{rodriguez2011nonparametric}. Choosing the probit stick-breaking process (PSBP) as prior, various models can be approximated while computational simplicity is preserved \cite{rodriguez2011nonparametric, chung2009nonparametric}. 

Following the single-atom characterization \cite{rodriguez2011nonparametric} of $G$, we can write it as an infinite mixture:
 \begin{equation}\label{mod: G}
 G(\cdot) = \sum_{k \geq 1} \omega_k\delta_{\gamma_k}(\cdot)
 \end{equation}
where $k$ indicates the $k$-th component of the infinite mixture and $\delta_{\gamma_k}$ is the Dirac measure at $\gamma_k$. $\{\gamma_k\}_{k \geq 1}$ and $\{\omega_k\}_{k \geq 1}$ are infinite sequences of the parameters and weights of the kernel, respectively, and are considered random variables. We assume $\{\gamma_k\}_{k \geq 1}$ and $\{\omega_k\}_{k \geq 1}$ comprise independent random variables and the sequences are independent of each other.
Moreover, following the stick breaking representation, the weights $\{\omega_k\}_{k \geq 1}$ are defined as
\[
\omega_k  =  u_k\prod_{r<k}(1-u_{r})
\]
where $\{u_k\}_{k>1}$ and $\{u_r\}_{r<k}$ are $[0, 1]$ valued.
The PBSP further assumes $u_k$, for each $k$, is a probit transformation of a random variable $\alpha_k$ that follows a Gaussian distribution with variance equal to 1:
\begin{equation} \label{eqn: u}
  u_k =  \Phi (\alpha_k), \quad \alpha_k \sim N(\mu_{\alpha}, 1)
\end{equation}
where $\Phi$ is a probit transformation of a random variable.

According to the discrete nature of $G$, we can introduce the latent categorical variable $S$, indicating the allocation to the components of the mixture. Specifically, assuming $Pr(S=k)= \omega_k$, the model \eqref{mod: Y} can then be written as:
\[
Y|\phi, S=k \sim f(\cdot | \phi_k)
\]
where $\phi_k$ are the specific parameters for the component of the mixture model $k$. 

\section*{Bayesian Nonparametric Approach}

\subsection*{Target Parameter and Assumptions for the True Model}

Our main interest is estimating the causal exposure-response function (CERF), defined as $E[Y(x)]$. Let $L$ be measured confounders, we assume: 
 \begin{itemize}
\item A1. consistency: $Y(x) = Y \mbox{ when } X = x$;
\item A2. Latent ignorability: $Y(x) \indep X|U,L$ for all $x$.
\end{itemize}

For convenience, we suppress the notation $L$ throughout the paper, but the results presented below are all conditional on covariates $L$, and our implementation of the following estimation procedure in {\it R} accommodates covariates $L$. Under assumptions A1 - A2, the CERF  can be expressed as
\begin{eqnarray*}
E[Y(x)] &= & E_U[E(Y(x)|X =x,  U)] \\
&=& E_U[E(Y|X =x,  U)]. 
\label{eqn: cf}
\end{eqnarray*} 
The notation $E_U[\cdot]$ is used to help clarify the conditional expectation is taken over $U$. 
\subsection*{Proposed Model and Assumptions}

\label{subsec:model}
Starting from Shi et. al's\cite{shi_selective_2020} idea, we relax one of their main assumptions that the outcome $Y$ is linearly related to the exposure $X$ and the unmeasured confounder $U$. We allow this relationship to be non-linear. We consider a mixture model for $Y$, using a linear function of $X$ and $U$ as the parametric kernel and a dependent PSBP as the prior. We introduce the latent categorical variable $S$, which indicates the data allocation to the mixture components, and we allow $S$ to depend on $X$. More specifically, we propose the following model.

 \begin{align}
  Y|X,U \sim \sum_{k \geq 1} \omega_k(X) f_k(Y|X, U)\notag; \\ 
  \omega_k(X)  =  \Phi(\alpha_k(X))\prod_{r<k}(1-\Phi(\alpha_r(X)).
  \label{model:y}
 \end{align}
In \eqref{model:y}, 
\[
Pr(S=k|X)= \omega_k(X)
\]
and $\alpha_k(X)$, for $k \geq 1$, is a Gaussian random variable that depends on the exposure variable $X$, such as
\begin{equation}
    \alpha_k(X) \sim N(\mu_{\alpha,k}(X),1).
    \label{eq:alpha_x}
\end{equation}
In this article, we study two models for $\mu_{\alpha,k}(X)$. In the first model, we assume $\mu_{\alpha,k}(X)$ is a continuous linear function of $X$:
\begin{equation}
    \mu_{\alpha,k}(X) = \eta_{0,k} +\eta_{1,k}X.  \label{weight_1}
\end{equation}

In the second model, we divide the observed domain for $X$ into 4-quantiles and assume $ \mu_{\alpha,k}(X)$ is a piecewise linear function of $X$. Let $q_1, q_2, q_3$ be the observed quartile cut-off points of $X$. We  assume
 \begin{eqnarray}\label{eqn: eta_4}
  \mu_{\alpha,k}(X)=& (\eta_{0,k} + \eta_{1,k}X)1(X \leq q_1)+ (\eta_{2,k} + \eta_{3,k}X)1(q_1 < X \leq q_2) \nonumber\\
  + &(\eta_{4,k} + \eta_{5,k}X)1(q_2 < X \leq q_3) + (\eta_{6,k} + \eta_{7,k}X)1(X > q_3) \label{weight_2}.
 \end{eqnarray}
Note our software supports even more flexibility for modeling $\mu_{\alpha,k}(X)$. Users can divide the $X$ domain into an arbitrary number of quantiles or at specified fixed points.

We further assume we can find auxiliary variables $Z$ and $W$ such that within each mixture component
\begin{itemize}
    \item A3. ($Z$ is an NCE ): $Z \indep Y|X, U, S=k$;
    \item A4. ($W$ is an NCO  ): $X \indep W| U, S=k$;
    \item A5. (NCE and NCO are conditionally independent): $Z \indep W|U, S=k$;
 \item A6.
\begin{eqnarray}
  Y|X,U, S=k  & \sim  & \mathcal{N}(\beta_{0,k}+\beta_{X,k}X  +\beta_{U,k}U,\sigma_{y,k}^2)  \label{eqn: f(y)};\\
    W| U, S=k   &\sim  & \mathcal{N}(\beta_{W0}+\beta_{WU}U,\sigma^2_w) \label{eqn: f(w)};\\
    U|X,Z, S=k    &\sim&   \mathcal{N}(\beta_{U0}+\beta_{UX}X + \beta_{UZ}Z,\sigma^2_u)\label{eqn: f(u)};
\end{eqnarray}  
    \item A7. (W is informative about U): $\beta_{WU} \neq 0 $; 
    \item A8. (Z is informative about U):  $\beta_{UZ} \neq 0$.
\end{itemize}
We can write the model \eqref{model:y} and \eqref{eqn: f(y)} conditioned on the latent categorical variable $S$ as the following:
\begin{equation}
    Y|X,U, S=k \sim f_k(Y|X, U)
    \label{mod:S}
\end{equation}
where $f_k$ is specified in \eqref{eqn: f(y)}. The definition of the model in \eqref{mod:S} clarifies the link between ours and the result of Shi, et.al \cite{shi_selective_2020}. By conditioning on each component of the mixture $k$ and assuming a linear model for $f_k(Y|X, U)$, we invoke the closed-form expression for the causal effect of $X$ on $Y$ for linear models. Using the mixture model in \eqref{model:y},  we can capture the non-linearity of the CERF that the linear model cannot. 

\subsection*{Approximation of  $E[Y(x)]$}

First, for each $k, k=1, 2, \dots, \infty$ we take conditional expectation on both sides of the equations \eqref{eqn: f(y)} and \eqref{eqn: f(w)} in A6 with respect to $f(U|X, Z)$. As a result,
\begin{eqnarray}
E_U[E(Y| X,Z, U, S=k) |X, Z]  &  = & \beta_{0,k}+\beta_{X,k}X  +\beta_{U, k}E(U|X,Z); \label{eqn: Y|X,Z,U)} \\
E_U[E(W|X,Z, U, S=k)|X,Z]   &  =  &  \beta_{W0}+\beta_{WU}E(U|X,Z) \label{eqn: W|U)}. 
\end{eqnarray}
Because $\beta_{WU} \neq 0 $ in A7, we have
\begin{equation} \label{eqn: modeled}
    E(Y| X,  Z, S=k) = \beta_{0,k}+\beta_{X,k}X  + \beta_{U, k}\frac{E(W|X, Z, S=k) - \beta_{U0}}{\beta_{WU}}.
\end{equation}
Next,  we reparameterize equations \eqref{eqn: Y|X,Z,U)} and \eqref{eqn: W|U)} and obtain
\begin{align}
    Y| X, Z, S=k &  \sim  \theta_{0, k}+\theta_{X,k}X  +\theta_{Z, k} Z +\delta_{y,k} \mathcal{N}(0,1); \label{eqn: Y|X,Z} \\
W|X, Z, S=k   & \sim   \theta_{W0} +\theta_{WX} X +\theta_{WZ} Z +\delta_w \mathcal{N}(0,1) \label{eqn: W|X,Z}. 
\end{align}
Comparing these two models to \eqref{eqn: modeled} yields
\begin{eqnarray}
\beta_{X,k} = \theta_{X,k}-\theta_{Z, k}\frac{\theta_{WX}}{\theta_{WZ}}.
\label{eqn: beta_k}
\end{eqnarray}
As a result, we can identify $\beta_{X,k}$ in \eqref{eqn: f(y)} that involves the unobserved variable $U$ by observed variables because the right-hand side of \eqref{eqn: beta_k} comprises parameters that can be estimated from the observed variables $Z, W, X, Y$ once the latent group membership $S$ is found by BNP algorithms. Note that the estimation of $\beta_{X,k}$ requires $\theta_{WZ} \neq 0$. This holds under assumptions A7 and A8 that $\beta_{WU}\neq 0$ and $\beta_{UZ} \neq 0$.

Although  $\beta_{U,k}$ and $E(U)$ are not identifiable,  we can identify the following under A6
\begin{equation}\label{eqn: intercept}
\beta_{0,k} + \beta_{U,k}E(U)  = \theta_{0, k} + \theta_{Z, k}E(Z) +\theta_{Z, k}\frac{\theta_{WX}}{\theta_{WZ}}E(X) .
\end{equation}
See proof in Appendix A.
Combining results in \eqref{eqn: beta_k} and \eqref{eqn: intercept}, we obtain
\begin{align}\label{eqn: y(x)}
     E[Y(x)] =& E_U[E(Y|X=x, U)]  \qquad \mbox{by A1 and A2} \nonumber\\
     = & E_U[E_S[E(Y|X=x, U, S)|X=x, U]] \nonumber\\
     = & E_U[P(S=k|X=x, U)E(Y|x, U, S)]\nonumber\\
     \approx & \sum_{k=1}^{\infty} E_U[P(S=k|X=x)(\beta_{0,k} + \beta_{X, k}x+\beta_{U,k}U )]\nonumber\\
    = &\sum_{k \geq 1} w_k(x)\beta_{X, k}x 
             + \sum_{k \geq 1} w_k(x)\left[\beta_{0, k} + \beta_{U,k}E_U(U)\right]  \nonumber\\
                         = & \sum_{k \geq 1} w_k(x)\left[\theta_{X,k}-\theta_{Z, k}\frac{\theta_{WX}}{\theta_{WZ}}\right]x \nonumber\\
            & + \sum_{k \geq 1} w_k(x)\left[\theta_{0, k} + \theta_{Z, k}E(Z) +\theta_{Z, k}\frac{\theta_{XW}}{\theta_{ZW}}E(X)\right] \qquad \mbox{by \eqref{eqn: beta_k} and \eqref{eqn: intercept}}.
\end{align}
The approximation occurs in two places. First, we consider that the mixtures of Gaussian linear models can well approximate the true model underlying $E[Y(x)]$. Second, since we can not observe $U$,  we approximate the true weight $P(S=k|X=x, U)$ by weight $P(S=k|X=x)$, relying on the considerable flexibility a dependent PSBP prior provides.  Under certain conditions,  the approximation sign $\approx$ can be replaced by $=$, and as a result, we can identify $E[Y(x)]$. We leave the investigation of such conditions for future work.

\subsection*{Estimation}

We estimate the posterior distributions of CERF defined in \eqref{eqn: y(x)}  using a Bayesian approach.  In Appendix B, we specify the conjugate priors that allow us
to conveniently obtain the posterior distributions for the parameters in \eqref{eqn: y(x)}. These include   1)  $\boldsymbol{\theta}_{Y,k} = [\theta_{0,k}, \theta_{X,k}, \theta_{Z, k}]$ and $\delta_{y,k}$ in \eqref{eqn: Y|X,Z}, which are the model parameters for  $Y|X, Z$;  2) $\boldsymbol{\theta}_W = [\theta_{W0}, \theta_{WX}, \theta_{WZ}]$ and $\delta_{w}$ in \eqref{eqn: W|X,Z},  which are the model parameters for $W|X, Z$, and 3)   ${\boldsymbol\eta}_k = [\eta_{0,k}, \eta_{1,k}, \eta_{2,k}, \dots, \eta_{p,k}]$, which  are the  parameters in the weight model \eqref{weight_1}  or \eqref{weight_2} for $\omega_k(x)$.  Rodrguez and Dunson \cite{rodriguez2011nonparametric} show that we can truncate the infinite mixture models in \eqref{eqn: y(x)} to a finite mixture with a reasonable conservative upper bond $K>0$ without changing the results. Thus, we consider $k = 1, 2, \dots, K$. In Appendix B we also provide the posterior distributions of these parameters. We then compute the posterior distribution for the approximate $E[Y(x)]$ by combining these distributions based on \eqref{eqn: y(x)} and propose to use the posterior median as our BNP-NC estimator $\hat{E}[Y(x)]$ for $E[Y(x)]$. The algorithm \ref{alg:model} in Appendix C outlines our Gibbs sampling procedure to estimate these parameters.  When using the weight prior model \eqref{weight_2}, our software provides the option of applying kernel smoothing to $\hat{E}[Y(x)]$.  Smoothing is implemented using \texttt{R} function \texttt{ksmooth} with the Gaussian kernel and the bandwidth of $0.2$. All analyses were performed using R Statistical Software (v4.4.0; R Core Team 2021)

\section*{Simulation Studies}
\label{sec:sim}

In this section, we present simulation studies to demonstrate that our method can recover the true shape of non-linear CERFs in the presence of an unmeasured confounder. We evaluate the performance of our method in four scenarios with varying nonlinear CERF shapes.   Let $n$ denote the sample size. In each of the four scenarios, for $i=1,\dots, n$, we first simulate $U_i$  following a normal distribution with mean $1$ and variance $0.3$ denoted by $U_i \sim \mathcal{N}(1,0.3)$. Next, we simulate the NCO variable $W_i \mid U_i \sim \mathcal{N}(1-2\cdot U_i,0.2)$ and the  NCE variable $Z_i\mid U_i \sim\mathcal{N}(-1+1.5\cdot U_i,0.2)$. Then, we simulate  $X_i$ and $Y_i$ under four scenarios that generate varying CERF shapes.
\begin{itemize}
    \item  {\bf Scenario 1}: the CERF follows a piecewise linear model:
\begin{align}
\begin{split}
     & X_i \mid U_i \sim \mathcal{N}(1.5+4\cdot U_i,0.5)\\
     & Y_i|X_i, U_i \sim \mathbb{I}_{\{X_i<5.5\}}\cdot \mathcal{N}(1+2\cdot X_i+2\cdot U_i, 0.3) \nonumber\\
   + &\mathbb{I}_{\{X_i\geq 5.5\}}\cdot \mathcal{N}(-16+5\cdot X_i+2.5\cdot U_i, 0.3)   
\end{split}
\end{align}
where $\mathbb{I}_{\{X \in \mathcal{A}\}}$ is an indicator variable, which takes the value of $1$ when the realization of the random variable $X$ belongs to the interval $\mathcal{A}$, and $0$ otherwise.

\item {\bf Scenario 2}:, the CERF is a parabola:
\begin{align}
\begin{split}
   & X_i \mid U_i \sim \mathcal{N}(2.5+4\cdot U_i,0.5) \\
    & Y_i|X_i, U_i \sim \mathcal{N}(-10+1.5\cdot(X_i-6)^2+4\cdot U_i,0.3).
\end{split}
\end{align}

\item {\bf Scenario 3}: the CERF has a sigmoidal shape:
\begin{align}
\begin{split}
    &  X_i \mid U_i \sim \mathcal{N}(1+4\cdot U_i,0.5)\\
    &Y_i|X_i, U_i \sim \mathcal{N}(1 / (1 + exp(-5 \cdot (X_i - 5)))+1.7\cdot U_i,0.1).
\end{split}
\end{align}  

\item {\bf Scenario 4}:, the CERF is monotonically increasing with a nonlinear relationship with both variables $X$ and $U$:
\begin{align}
\begin{split}
    & X_i \mid U_i \sim \mathcal{N}(2.5 +4\cdot U_i,0.5)\\
    &Y_i \sim \mathcal{N}(-2\cdot e^{-1.4\cdot (X_i-6)}+1.5\cdot e^{U_i},0.3).
\end{split}
\end{align} 
\end{itemize}
We set $n=5000$.  To preserve the characteristic of the normal distribution while avoiding data scarcity for extreme values, we simulate $6250$ data points and truncate the data in the 10th and 90th percentile of $X_i$ to obtain the sample size of $5000$.

\begin{figure*}[tbp!]
\begin{center}
\includegraphics[width=2.6in]{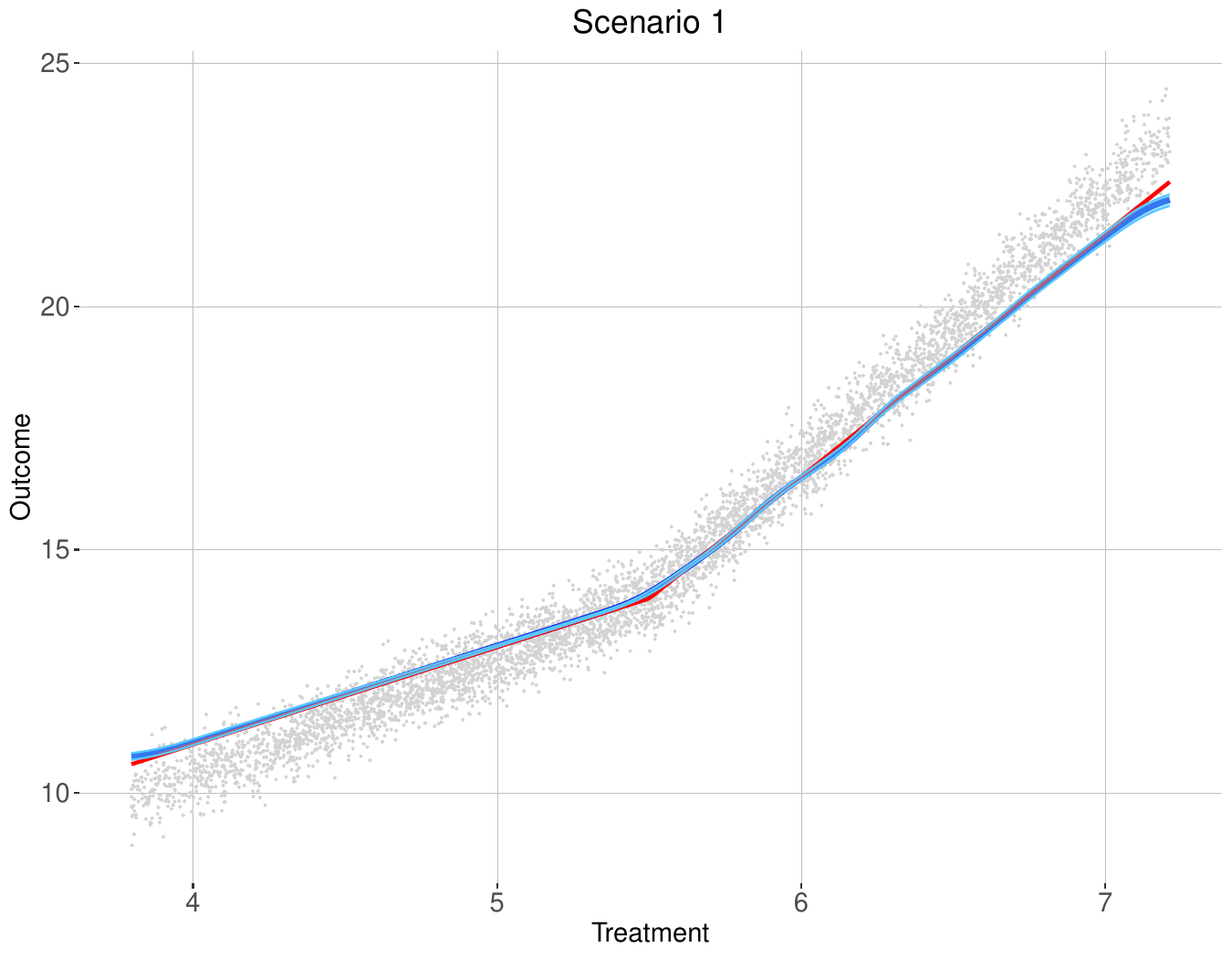}\;\includegraphics[width=2.6in]{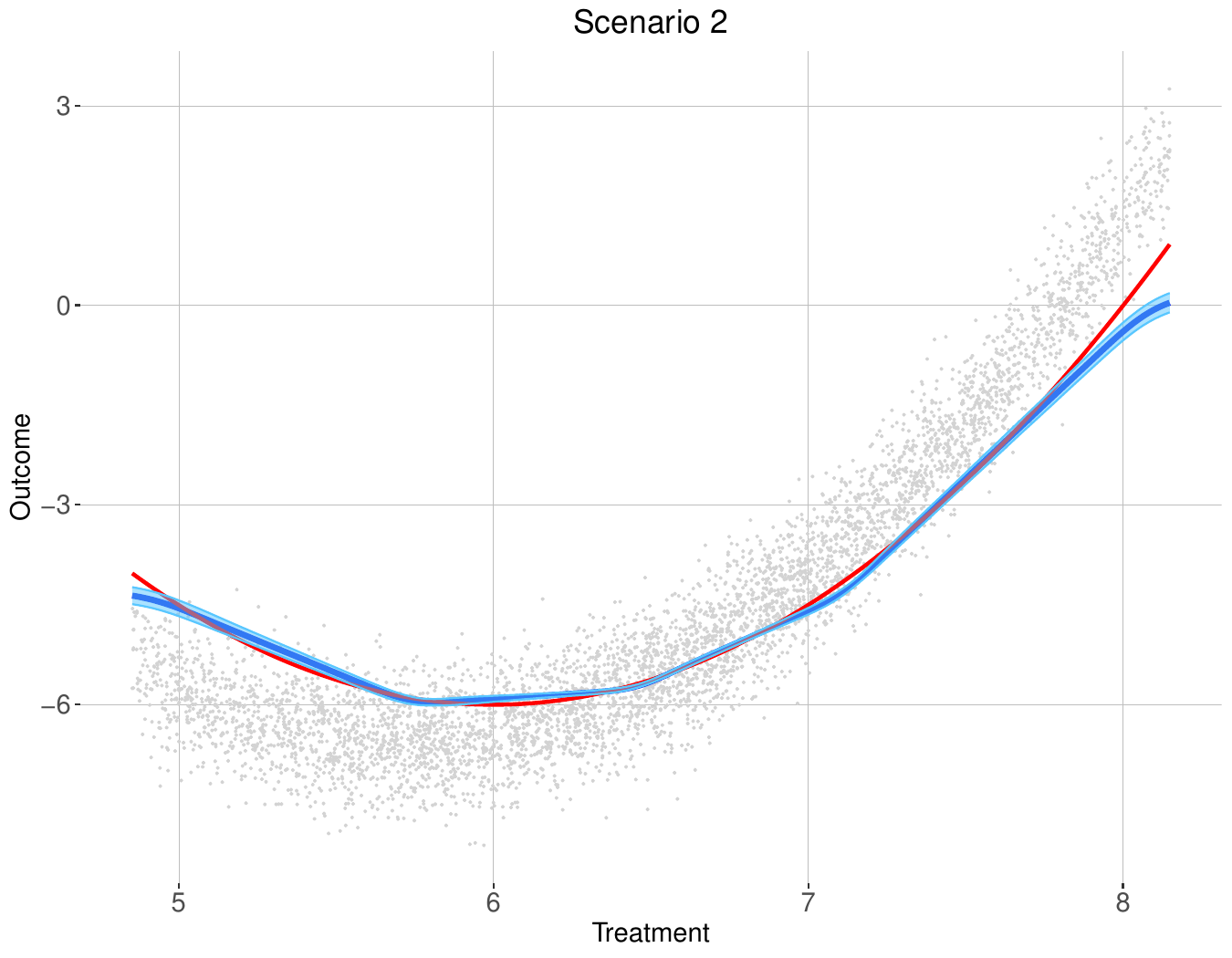}\\
\includegraphics[width=2.6in]{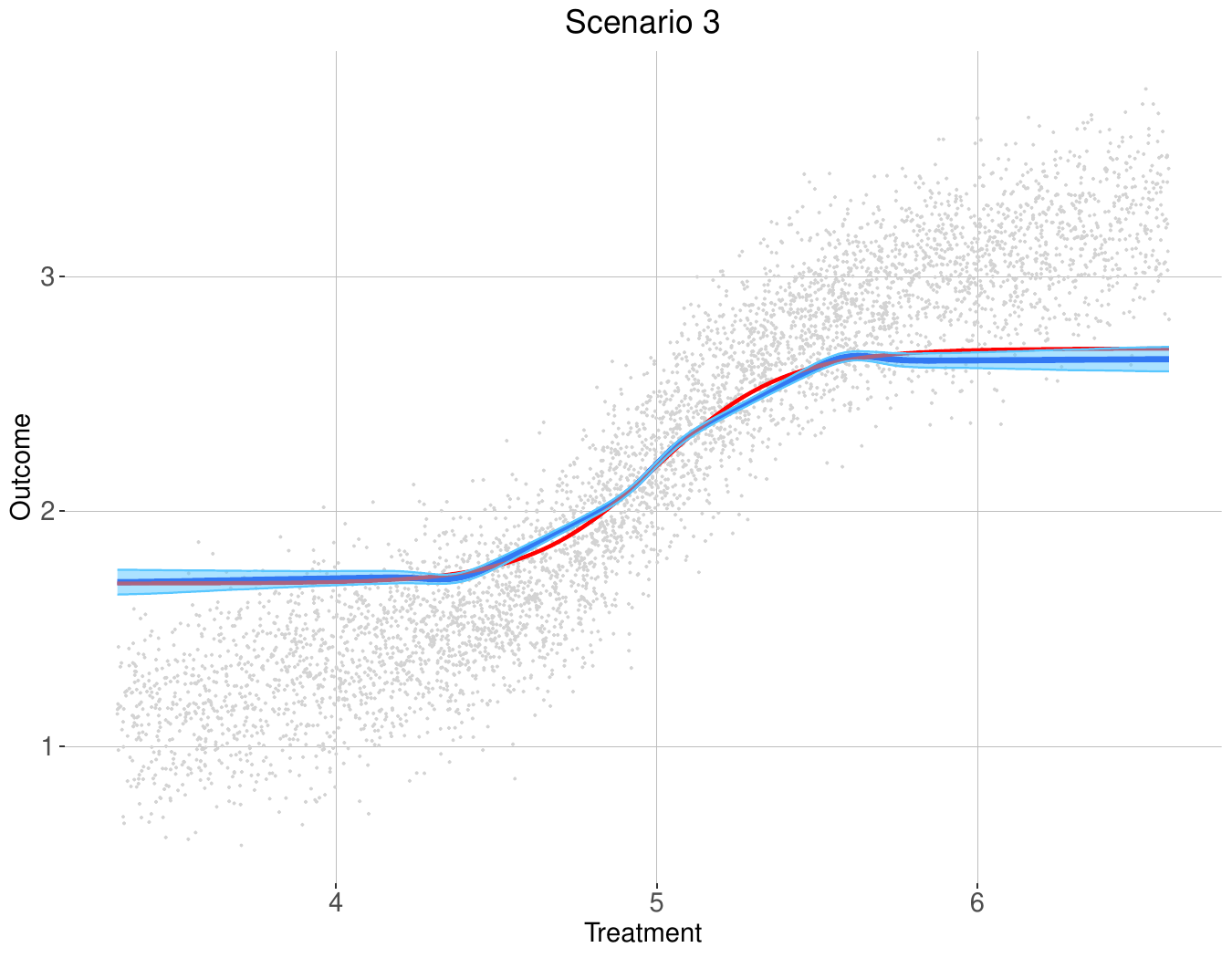}\;\includegraphics[width=2.6in]{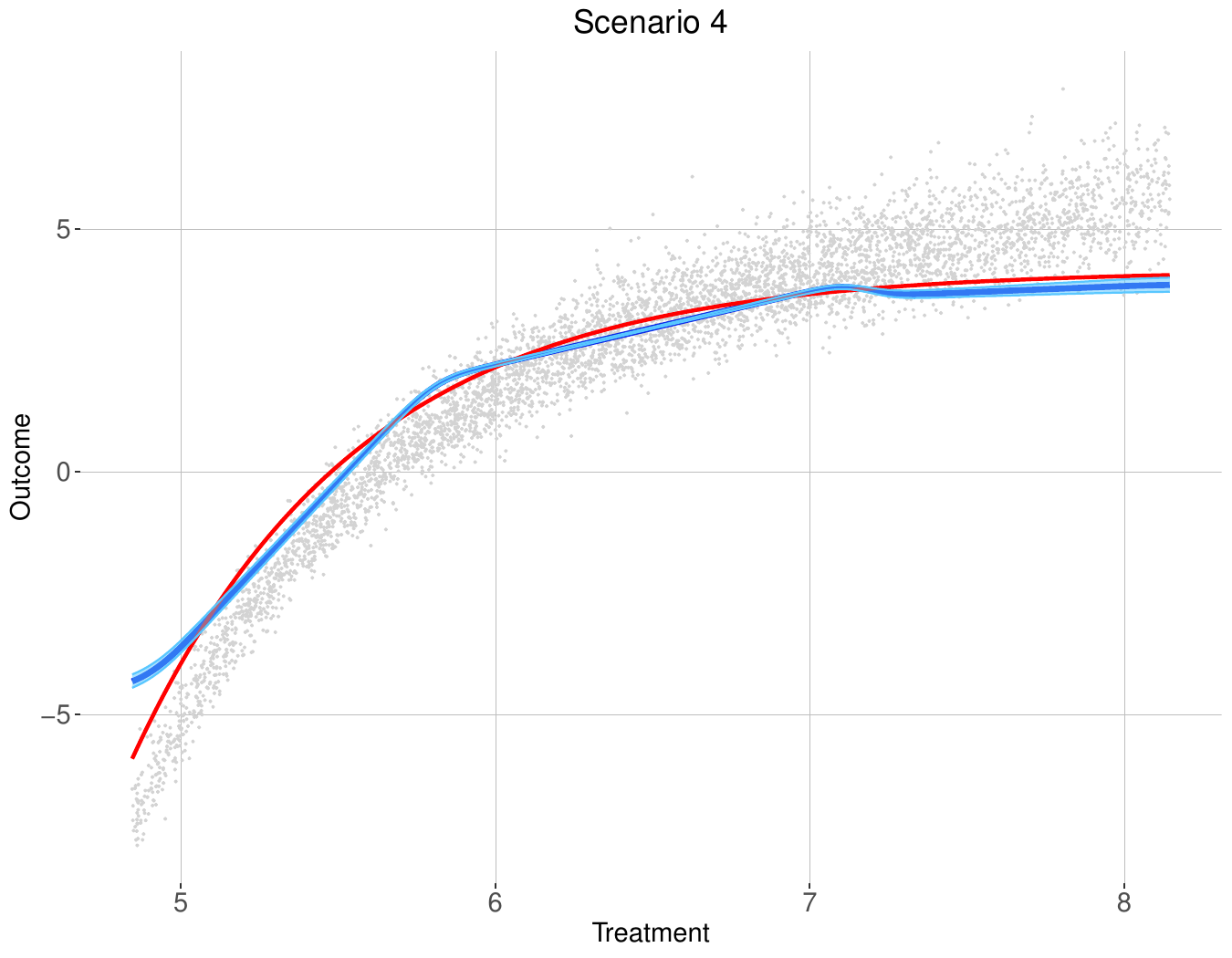}
\end{center}\caption{Comparison of the estimated CERFs (the darker blue line) with 95\% credible intervals (the light blue band) to  the true CERFs (the red line) of varying shapes.  Light gray dots show the simulated data. The sample size is $5000$. The prior for the mixture weights follows model \eqref{weight_2}.    \label{fig:results_4quantiles}}
\end{figure*}

In Figure~\ref{fig:results_4quantiles}, we show the fitted CERF when using model  \eqref{weight_2} to construct the prior for the mixture weights. Figure~\ref{fig:no_smoothing} in Appendix D shows the results before smoothing and Figure~\ref{fig:results_4quantiles} shows the ones after smoothing.
The  red line in Figure~\ref{fig:results_4quantiles} is the true CERF. Light gray dots are the simulated data,  indicating the relationship between $X$ and $Y$ without adjusting for $U$. The discrepancy between the data points in gray and the true CERF in red suggests the presence of unmeasured confounders. The darker blue line is the posterior median of the CERF estimates in the last $1000$ of $2000$ Gibbs sampling iterations. Figure~\ref{fig:results_4quantiles} and Figure~\ref{fig:no_smoothing} show the estimated CERFs closely follow the true CERFs(red curves)  throughout most of the treatment distribution's support in these four scenarios. Without smoothing, Figure~\ref{fig:no_smoothing} shows the estimated CERFs are discontinuous due to discontinuity of the model \eqref{weight_2} that defines the prior for the mixture weights.

We perform a sensitivity analysis on the choice of the model $\mu_{\alpha, k} (X)$ in \eqref{eqn: eta_4} that defines the mixture weights.  We consider an alternative model \eqref{weight_1} proposed by Rodriguez and Dunson \cite{rodriguez2011nonparametric}, which assumes $\mu_{\alpha, k} (X)$ follows  a linear regression function of $X$ without dividing the support of $X$ into quantiles. Figure \ref{fig:no_splitting} in Appendix D shows the 95\% CIs of the estimated CERFs (the light blue band) cover the true CERFs in most parts of the support of $X$ in all scenarios. However, the discrepancy between the estimated CERFs and the true CERFs are wider than those using  model \eqref{weight_2},  possibly because the weight model \eqref{weight_1} is not flexible enough.  Although model \eqref{weight_1} gives a smooth CERF estimate without the additional step of kernel smoothing, its lack of flexibility for certain  CERF shapes motivates us to adopt weight \eqref{weight_2} for real-data applications in the next section.

Despite the slight deviation, the simulation studies demonstrate that the proposed method with the weight prior \eqref{weight_2} is very flexible in different settings when the non-linear CERFs are of various shapes and the sample size is moderate at 5000. The estimated CERFs for all scenarios follow the true CERFs rather than the apparent but distorted relationship between $X$ and $Y$ in gray due to the unmeasured confounder $U$,  demonstrating the effectiveness and utility of our method to adjust for unmeasured confounding.
 
\section*{Application: long-term exposure to ambient $PM_{2.5}$ and cardiovascular hospitalization rates}
In this section, we illustrate how to select NC in practice and use our method to estimate the CERF of long-term exposure to ambient $PM_{2.5}$, on cardiovascular hospitalization rates among the elderly in the continental US \cite{papadogeorgou2020causal}. 

Our analysis is at the zip code level, labeled $i$ , and includes $N =5362$ zip codes.  The outcome $Y_i$ is defined as the logarithm of the hospitalization rate for cardiovascular diseases (codes ICD-9 390 to 459) among Medicare beneficiaries residing in zip code $i$ during 2013. Exposure $X_i$ is defined as the average daily levels of ambient concentrations $PM_{2.5}$ for 2011 and 2012 recorded by EPA (U.S. Environmental Protection Agency) monitors within a 6-mile radius of the centroid of zip code $i$. The data set does not cover all zip codes in the continental United States, but those within this radius of an EPA monitoring site with more than 67\% scheduled measurements.  In addition, a large collection of covariates was assembled at the zip code level, including information on socioeconomic, demographic, and cardiovascular disease risk factors.  See Table A.1. in \cite{papadogeorgou2020causal} for detailed information on these covariates.

For illustration purposes, we consider one measured covariate, the logarithm of median household income, as an unmeasured confounder ($U$), and the CERF benchmark is the estimated CERF after adjustment for $U$.  To examine the performance of our approach, we mask the variable $U$ and compare whether using our BNP method with NC can recover the CERF benchmark. We acknowledge that the CERF benchmark is not the true CERF, which is unknown for real data applications, and we can further improve the benchmark CERF by adjusting for measured confounders that we do not consider in this example. Hence, the reader should not view our results as scientific evidence to quantify the causal effect of long-term exposure to ambient $PM_{2.5}$ on cardiovascular hospitalization rates among the elderly in the continental United States, but as an illustration of novel statistical methods to address unmeasured confounding bias.

As depicted by the causal diagram in Figure \ref{fig: pm}, the household income can be a potential confounder because low-income people are more likely to live in low-cost areas, which could be associated with an elevated level of $PM_{2.5}$ (X), for example areas near highways.  Compared to a zip code with a higher household income, low-income people  may face challenges to afford health care. As a result, a low household income for a zip code can be associated with a low hospitalization rate due to cardiovascular diseases $(Y)$.  Without accounting for the average household income,  Evaluating the impact of ambient $PM_{2.5}$ on cardiovascular hospitalization rates may lead to biased results if the median household income is not taken into account.

\begin{figure}[tbp]
    \hspace{-0.5cm}
\begin{tikzpicture}
    \node[state] (1) [xshift = 2cm]{$X$: $PM_{2.5}$};
    \node[state] (2) [right =of 1, align=left] {$Y$: log \\ hospitalization rate};
    \node[state] (3) [above =of 1, yshift = -1.5cm, xshift = 2.5cm] {$U_k$: log median income};
    \node[state] (4) [above =of 1, xshift = 1.5cm, align=left] {$Z$: \% employed \\ population};
    \node[state] (5) [right =of 4, xshift = -1.3cm,  align=left] {$W$: \%  owner \\ occupied houses };
    \path (1) edge node[below] {} (2);
    \path (3) edge node[below] {} (1);
    \path (3) edge node[above] {} (2);
    \path[] (4) edge node[above] {} (3);
    \path[] (3) edge node[below] {} (5);
\end{tikzpicture}
    \caption{The causal diagram of the unmeasured confounder $U$, the negative control exposure $Z$, and the negative control outcome $W$.}
    \label{fig: pm}
 \end{figure}

One crucial step to use our approach to address unmeasured confounders is to select the appropriate NC. In this analysis, we consider the proportion of residents in a zip code who have jobs as our NCE variable $Z$. We consider the percentage of owner-occupied housing units in each zip code as our NCO variable $W$.  Conditional on household income $(U)$, we make the strong assumption that, for each zip code, the ambient  $PM_{2.5}$ level $(X)$ is independent of the percentage of owner-occupied housing units $(W)$,  the percentage of employment $(Z)$ is independent of the hospitalization rate for cardiovascular diseases among the elderly $(Y)$, and the percentage of employment $(Z)$ and the percentage of owner-occupied housing units $(W)$ are also independent (See Figure \ref{fig: pm}). We recognize that these assumptions might not hold and are not testable in practice. However, for this illustrative example, we can check whether these assumptions hold for our data since $U$ is known. We use conditional independence tests in \texttt{R} package \texttt{dagitty} \cite{dagitty} developed for continuous data to verify our assumptions. We check whether we can reject the null hypothesis $H_0$ by verifying whether the 95\% confidence interval covers $0$ for each row of Table 1 and conclude the conditional independence assumptions in A3, A5, A7, and A8 hold and this dataset does not violate Assumption A4 much. 

\begin{table*}[tb]
  \centering
  \caption{Hypothesis Testing of Assumptions A3-A7}
  \label{tab:conditional_independence}
  \begin{tabular}{|c|c|c|c|c|}
    \hline
  Assumptions & $H_0$ & Estimate & 95\% CI & Conclusions\\
     \hline
A3*  & $ Z \perp Y |X, U$  &   0.0049  & [-0.0288, 0.0374]& A3 holds \\    
 A4*& $ X \perp W| U$ & -0.0368 & [-0.0619, -0.0119] & A4 almost holds \\    
A5*& $W \perp Z |U$ & -0.02172  & [-0.0644, 0.0238] & A5 holds \\    
A7: $\beta_{WU} \neq 0$ &  $W \perp U $ &0.3727 & [0.3284, 0.4173] & A6 holds\\
 A8: $\beta_{UZ} \neq 0$ & $U \perp Z |X $ & 0.5792 & [0.5571, 0.6009] & A7 holds\\
 \hline
\multicolumn{5}{l}{\footnotesize A3*-A5*: Here we test marginal conditional independence, slightly different from A3-A5}\\
\multicolumn{5}{l}{\footnotesize $X$: $PM_{2.5}$; $Y$: the logarithm of the hospitalization rate for cardiovascular diseases}\\
\multicolumn{5}{l}{\footnotesize $U$: the logarithm  of the median household income}\\
\multicolumn{5}{l}{\footnotesize $Z$: the percentage of the occupied population; $W$: the percentage of owner-occupied housing units }\\
\multicolumn{5}{l}{\footnotesize All the hypothesis tests are two-sides conducted at the significance level of 0.05}
\end{tabular}
\end{table*}

We fit three models to the data for comparison. The first model, denoted as ``YX" in Figure \ref{fig: results_data}, does not consider the unmeasured confounder $U$ and assumes 
\[
Y | X \sim \sum _{k \geq 1} w_k(X)\mathcal{N}(\beta_{0,k} + \beta_{X,k}X).
\]
In the second model denoted as ``YXU" in Figure \ref{fig: results_data}, we unmask $U$  and fit the BNP model, adjusting for $U$ by treating it as a covariate. $U$ is allowed to have a non-linear relationship with $Y$:
\[ 
Y | X, U \sim \sum _{k \geq 1} w_k(X)\mathcal{N}(\beta_{0,k} + \beta_{X,k}X +\beta_{u,k}U ).
\]
The third model, denoted as ``BNP-NC", accounts for $U$ without using $U$. We apply our BNP method described in Section 3, incorporating NC to adjust for $U$. 

 Figure \ref{fig: results_data} shows the estimated response functions using these three models and the 95\% CI of the estimate curves.  We consider the response function estimated from the ``YXU" model as our benchmark and use it to evaluate the performance of our proposed method. When the covariate $U$ is masked, the estimated response curve from the ``BNP-NC" model (the blue line and interval) is closer to the benchmark ``YXU" result (the green line and interval) than the response function estimated from the ``YX" model (the orange line and interval),
demonstrating the effectiveness of our method.  The discrepancy between the ``YX" modeling result (the orange line and interval) and the benchmark ``YXU" result (the green line and interval) highlights the danger of not accounting for the unmeasured confounder $U$. The small difference between the ``BNP-NC"  and ``YXU" results hints our NC possibly also adjusts for other unmeasured confounders than the household income $U$.
 


\begin{figure}[t]
\begin{center}
      \includegraphics[width=3.5in]{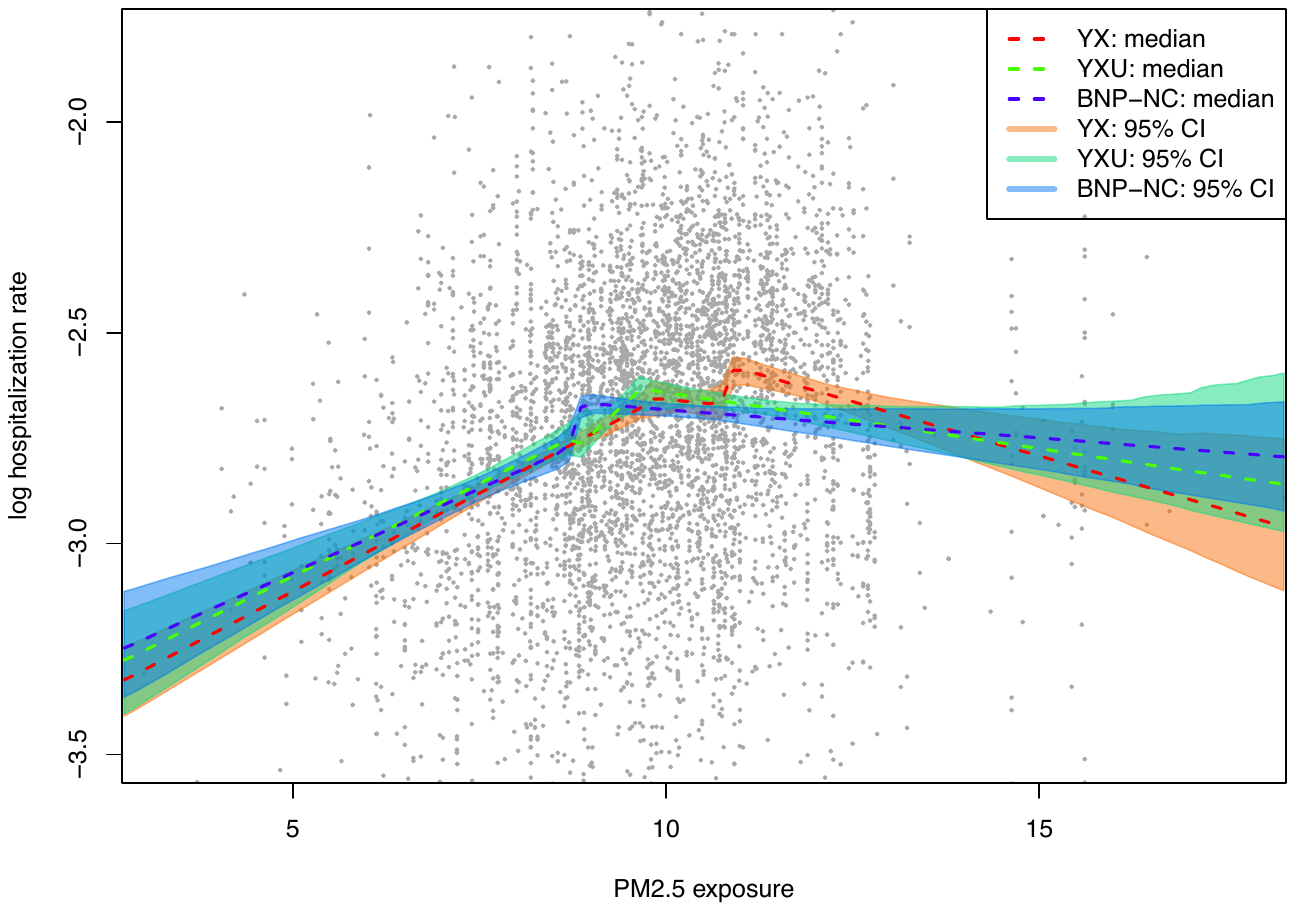}
\end{center}
\caption{The estimated response functions (dotted lines) and their 95\% credible intervals (colored areas) based on three models.``YXU", colored in green, indicates the result from the BNP method adjusting for a confounder $U$ when $U$ is unmasked. ``YX", colored in orange, indicates the result from the BNP method without considering the confounder $U$.``BNP-NC", colored in green, indicates the result from the BNP method using NC to adjust for unmeasured confounders when $U$ is masked. Grey dots are observed data points.\label{fig: results_data}}
\end{figure}

\section*{Discussion}
We develop a method to estimate the CERF indicated by $E[Y(x)]$ in the presence of  unmeasured confounders $U$ for the continuous exposure $X$ and the continuous outcome $Y$.  We address the following three challenges. 
\begin{enumerate}
    \item $U$ is a latent variable that standard causal inference techniques for confounding adjustment using either matching or inverse probability weighting do not apply. We solve this problem by using NC variables $Z$ and  $W$, which are often easy to obtain and sometimes already available among the collected covariates, as shown in our illustration with real data. 
    \item The response function $Y$ to a continuous variable $X$ is often non-linear and varies in different applications. Using a parametric model to capture all possible shapes of a response function is difficult. Our approach is to adopt a Bayesian nonparametric mixture model which can handle an arbitrary response curve shape.
    \item There is currently no method to incorporate NC (or proxy variables) into a Bayesian nonparametric estimation method to address unmeasured confounding issues. We fill this methodology gap by modeling the causal response function of $Y$ to $X$ of an arbitrary shape using a mixture of linear models and  leveraging a proximal causal inference result for linear models  when estimating the causal effect in the presence of $U$. 
\end{enumerate}
Our simulation studies have shown that our method can recover the true CERF of various shapes with finite samples under unmeasured confounding, providing an alternative solution to machine learning approaches that may require a large sample size to perform well. We illustrate how to use our method by applying it to studying the relationship between long-term exposure to ambient $PM_{2.5}$ on cardiovascular hospitalizations among the elderly in the continental US, showing the difference between using and without using our method and highlighting the danger of not accounting for unmeasured confounders.

There are several limitations in our methodology for future improvement. 
First, although we allow the confounding effect of $U$ on the result $Y$ to be non-linear in our model, we assume a linear relationship between $U$ and the NCO variable $W$, and between $U$ and the NCE variable $Z$. Such restrictions can be relaxed by using BNP models to define these relationships as how we model the outcome $Y$. Second, we assume that the conditional independence conditions in A3-A5 hold for each mixture component. However, our method performs well in the real-data example, even when A3-A5 only marginally holds. It is of theoretical interest to explore whether assumptions A3-A5 can be relaxed to a marginal assumption when data in each mixture follow Gaussian distributions. Third, we construct the weights of the mixture model such that they depend on $X$ and further allow the relationship of this dependence to vary in different quantiles of $X$ in \eqref{weight_2}. Comparing the simulation results from using models \eqref{weight_2} to \eqref{weight_1}, which restrict the relationship of this dependence the same for the entire support of $X$, we observe that although the estimated CERFs from using model \eqref{weight_2} are less smooth, they have a better performance for a moderate sample for all four scenarios investigated. Because our BNP approach allows any function of $X$ to define the mixture weights in \eqref{weight_2}, it would be interesting to explore other functional forms for the weight prior that can balance the smoothness and flexibility.

For future investigation and extension, our simulation results show that our estimators based on the approximation formula \eqref{eqn: y(x)} for $E[Y(x)]$  closely follow the true CERF in the four scenarios we studied. Theoretically, it would be interesting to study under what conditions our proposed formula identifies $E[Y (x)]$ and our estimators are consistent. When defining the weights for the mixture model, we assume the hyperparameter of the variance in \eqref{eq:alpha_x} equal to $1$  for all the components for computational efficiency reason. We can consider other values for this hyperparameter as well.  In simulation studies, more experiments can be designed to investigate how the proposed method performs when the NC variables $Z$ and $W$ become less correlated with $U$, the outcome data become noisier and the sample size varies. In addition, we find with a small sample size the model performance is sensitive to the random seed number and the maximum number of components we set at the start of the algorithm, due to the difficulty of estimating many parameters with few observed data. More investigation is warranted.  To increase the flexibility of the model, we let the weights of the mixture model depend on $X$. However, this method can also easily accommodate weight dependence on NC variables, which may further increase the method's performance.


In conclusion, our proposed method, which combines Bayesian nonparametric methods with the proximal causal inference framework, offers a powerful tool for estimating the CERF in the presence of unmeasured confounding. Our simulation studies demonstrate our method's effectiveness with finite sample sizes, and our real-data application example showcases how to use our tool in practice to address a common problem in observational studies.


\begin{figure}[t]
\begin{center}
      \includegraphics[width=3.5in]{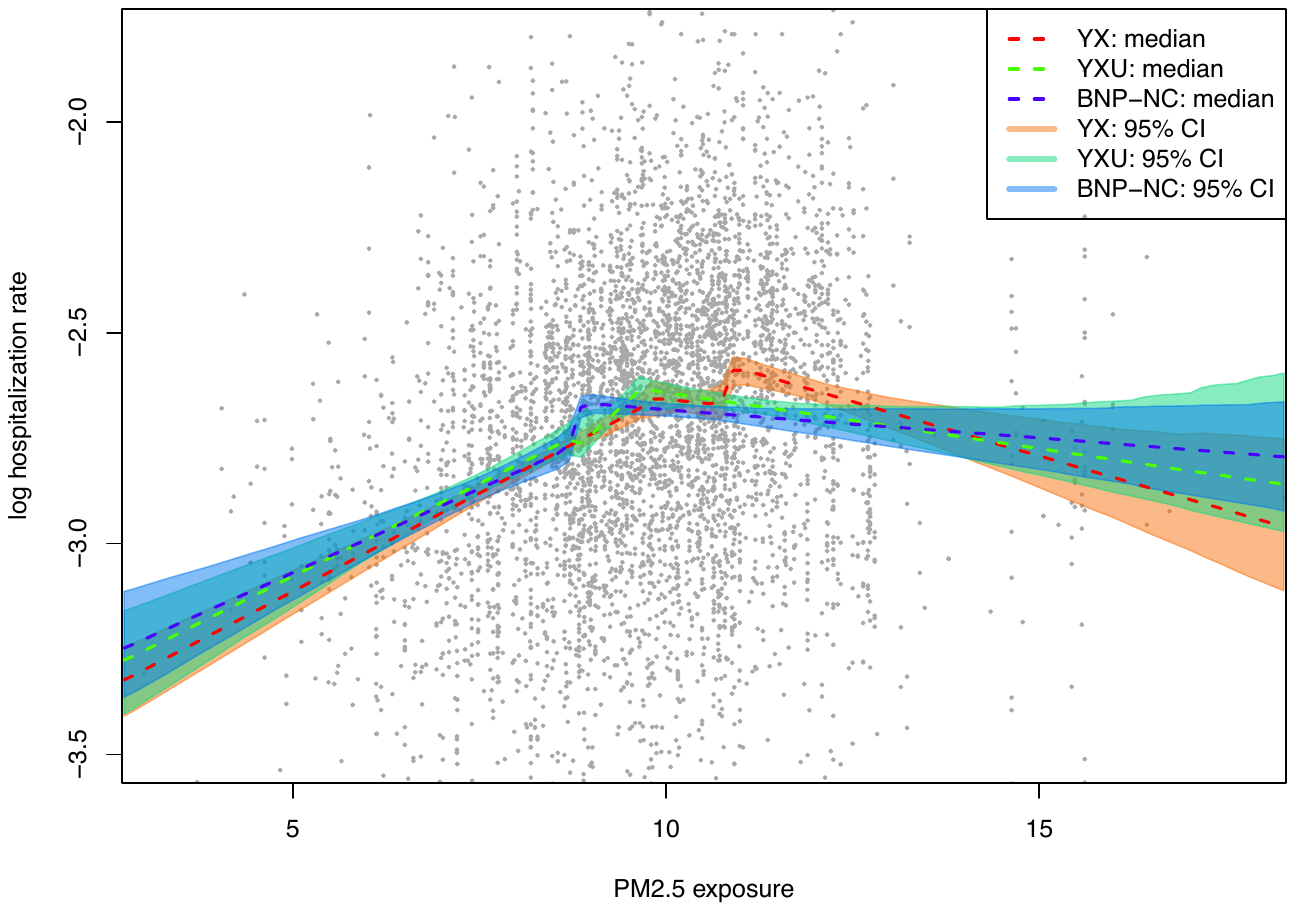}
\end{center}
\caption{The estimated response functions (dotted lines) and their 95\% credible intervals (colored areas) based on three models.``YXU", colored in green, indicates the result from the BNP method adjusting for a confounder $U$ when $U$ is unmasked. ``YX", colored in orange, indicates the result from the BNP method without considering the confounder $U$.``BNP-NC", colored in green, indicates the result from the BNP method using NC to adjust for unmeasured confounders when $U$ is masked.\label{fig: results_data}}
\end{figure}

\section*{Discussion}
We develop a method to estimate the CERF indicated by $E[Y(x)]$ in the presence of an unmeasured confounder $U$ for continuous exposure $X$ and continuous result $Y$.  We address three challenges in this causal inference problem summarized below. 
\begin{enumerate}
    \item $U$ is a latent variable that standard causal inference techniques for confounding adjustment using either matching or inverse probability weighting do not apply. We solve this problem by using a pair of auxiliary variables,  $Z$ and  $W$, which are often easy to obtain and sometimes already available among the collected covariates, as shown in our illustration with real data. 
    \item The response function $Y$ to a continuous variable $X$ is often non-linear and varies in different applications. Using a parametric model to capture all possible shapes of a response function is difficult. Our approach is to adopt a Bayesian nonparametric mixture model which can handle an arbitrary response curve shape.
    \item There is currently no method to incorporate NC (or proxy variables) into a Bayesian nonparametric method to account for unmeasured confounding. We fill this methodology gap by modeling the response of $Y$' to $X$ of an arbitrary shape 1) using a mixture of linear models and 2) leveraging a proximal causal inference result for linear models that provides a closed form solution when estimating the causal effect in the presence of $U$. 
\end{enumerate}
Our simulation studies have shown that our method can recover the true CERF of various shapes with finite samples in the presence of unmeasured confounding, providing an alternative solution to machine learning approaches that may require a large sample size to perform well. We illustrate how to use our method by applying it to studying the relationship between long-term exposure to ambient $PM_{2.5}$ on cardiovascular hospitalizations among the elderly in the continental US, showing the difference between using and without using our method and highlighting the danger of not accounting for unmeasured confounders.

There are several limitations in our methodology for future improvement. 
First, although we allow the confounding effect of $U$ on the result $Y$ to be non-linear in our model, we assume a linear relationship between $U$ and the NCO variable $W$, and between $U$ and the NCE variable $Z$. Such restrictions can be relaxed by using BNP models to define these relationships as we model the outcome $Y$. Second, we assume that the conditional independence condition in A3-A5 holds for each mixture component. However, our method performs well in the real-data example, even when A3-A5 only marginally holds for variables $Y$, $X$, and $Z$. It is of theoretical interest to explore whether assumptions A3-A5 can be relaxed to a marginal assumption when data follow Gaussian distributions. Third, we construct the weights of the mixture model such that they depend on $X$ and further allow the relationship of this dependence to vary in different quantiles of $X$ in \eqref{weight_2}. Comparing the simulation results from using models \eqref{weight_2} versus \eqref{weight_1}, which restrict the relationship of this dependence the same for the entire support of $X$, we observe that although the estimated CERFs from using model \eqref{weight_2} are less smooth, they have a better performance for a moderate sample for all four scenarios investigated. Because our BNP approach allows any function of $X$ to define the mixture weights in \eqref{weight_2}, it would be interesting to explore other functional forms for the weight prior that can balance the smoothness and flexibility.

For future investigation and extension, our simulation results show that our estimators based on the approximation formula \eqref{eqn: y(x)} for $E[Y(x)]$  closely follow the CERF in the four scenarios we studied. Theoretically, it would be interesting to study under what conditions our proposed formula identifies $E[Y (x)]$ and our estimators are consistent. When defining the weights for the mixture model, we assume the constant variance of $1$ in \eqref{eq:alpha_x}. We can consider other values for this variance as well.  In simulation studies, 
more experiments can be designed to investigate how the proposed method performs when the NC variables $Z$ and $W$ become less correlated with $U$, the outcome data become noisier and the sample size varies. In addition, we find from our simulation studies that with a moderate sample size the model performance is sensitive to the random seed number and the maximum number of clusters we set at the beginning of the algorithm. More investigation is warranted.  To increase the flexibility of the model, we let the weights of the mixture model depend on $X$. However, this method can also easily accommodate weight dependence on NCs, which may further potentially increase the performance in the tails of the response functions.


In conclusion, our proposed method, which combines Bayesian nonparametric methods with the proximal causal inference framework, offers a powerful tool for estimating the CERF in the presence of unmeasured confounding. Our simulation studies demonstrate our method's effectiveness with finite sample sizes, and our real-data application example showcases its practical utility in addressing a common problem in observational studies.

\section*{Acknowledgement}

 This work was supported by the National Institute Environmental Health Sciences grant (T32 ES 7069), Sloan Foundation grant (G-2020-13946), and the National Institute Health grants (R01ES028033, 1R01ES030616, 1R01ES029950, 
R01MD012769, 1RF1AG074372-01A1, 1RF1AG080948, 
1RF1AG071024).
\bibliographystyle{myama}
\bibliography{ref}

\newpage
\pagebreak
\appendix
\counterwithin{figure}{section}
\counterwithin{table}{section}
\pagenumbering{arabic}
    \setcounter{page}{1}
    
\begin{center}
    \large \textbf{SUPPLEMENTARY MATERIAL}
\end{center}

\input{derivation}
\section{Priors, Posteriors, and Gibbs Sampling}

\input{Posteriors}

\newpage
\section{Gibbs Sampling Algorithm}
\input{Gibbs}
\newpage
\section{Additional Simulation Studies}
\input{Additional_simulations}



\end{document}

%% file: derivation.tex
\section{Proof of Equation (16)}
\begin{align*}
& \beta_{0,k} + \beta_{U,k}E(U) \nonumber\\
= & E_{X,Z}[\beta_{0,k} + \beta_{U,k}E(U|X,Z)] \nonumber  \\
= & E_{X,Z}[\beta_{0,k} + \beta_{U,k}E(U|X,Z,S=k)] \qquad \mbox{ by \eqref{eqn: f(u)} in A6} \nonumber\\
=& E_{X,Z}[E(Y|X,Z, S=k) -\beta_{X,k}X] \qquad  \mbox{ by \eqref{eqn: f(y)} in A6} \nonumber\\
 = & E_{X,Z}[\theta_{0, k}+\theta_{X,k}X  +\theta_{Z, k}Z  -\beta_{X,k}X] \qquad 
 \mbox{ by \eqref{eqn: Y|X,Z}} \nonumber\\
 = & \theta_{0, k} + \theta_{Z, k}E(Z) +\theta_{Z, k}\frac{\theta_{WX}}{\theta_{WZ}}E(X) \qquad \mbox{by \eqref{eqn: beta_k}}.
\end{align*}

%% file: Posteriors.tex
In the following, we describe our Bayesian inference framework. First, we complete the definition of the model in Section~\ref{subsec:model} with the clarification of the prior distributions for the parameters. 

Let $\mathcal{N}_c(a,b\cdot\mathbf{1}_c)$ denote a $c$-variate Gaussian distribution with the mean vector $a$ and the covariance matrix $b\cdot\mathbf{1}_c$, i.e., a $c\times c$ diagonal matrix. With this notation, we assume independence among components of the vector and all components have the same variance $b$.

We assume 
the priors for ${\boldsymbol\eta}_k$ in the weight model follow:

\begin{equation}
   {\boldsymbol\eta}_k \stackrel{iid}{\sim} \mathcal{N}_p (\mu_{\eta,k},\sigma_{\eta}^2\cdot \mathbf{1}_p), \mbox{ for } k= 1, 2, \dots, K.  
   \label{eq:prior1}
\end{equation}
where ${\boldsymbol\eta}_k$ is the $p$-dimensional parameter vector in the probit regression model for the weights of the mixture model.

We assume the priors for the parameters ${\boldsymbol\theta}_{Y,k}$ and  $\delta_{y,k}$ in model \eqref{eqn: Y|X,Z} for $Y|X,Z, S=k$ follow 
\begin{align}
    {\boldsymbol\theta}_{Y,k} & \stackrel{iid}{\sim} \mathcal{N}_3(\mu_{\theta},\sigma_{\theta}^2\cdot \mathbf{1}_3), \mbox{ for } 
k= 1, 2, \dots, K; \notag\\
    \delta_{y,k} & \stackrel{iid}{\sim} \mbox{InvGamma}(\gamma_1,\gamma_2), \mbox{ for } 
k= 1, 2, \dots, K;
    \label{eq:prior2}
\end{align}
where ${\boldsymbol\theta}_{Y,k}=(\theta_{0,k},\theta_{X,k},\theta_{Z,k})$ are the component-specific regression parameters in the kernels of the mixture model, $\delta_{y,k}$ is the component-specific variance parameter for the kernels. We assume the prior for $\delta_{y,k}$ follows an inverse-gamma distribution with the shape parameter $\gamma_1$ and the scale parameter $\gamma_2$, with mean equal to $\frac{\gamma_2}{\gamma_1-1}$ and variance $\frac{\gamma_2^2}{(\gamma_1-1)^2(\gamma_1-2)}$.

We assume the priors for ${\boldsymbol\theta}_W$ and $\delta_{w}$ in the model \eqref{eqn: W|X,Z} for $W|X,Z, S=k$ distribution follow:
\begin{align}
    {\boldsymbol\theta}_W & \sim \mathcal{N}_3(\mu_{\theta},\sigma_{\theta}^2\cdot \mathbf{1}_3); \notag\\
    \delta_{w} & \sim \mbox{InvGamma}(\gamma_3,\gamma_4),
    \label{eq:prior3}
\end{align}
where ${\boldsymbol\theta}_W=(\theta_{W0},\theta_{WX},\theta_{WZ})$ are the regression parameters in the model \eqref{eqn: W|X,Z} for $W|X,Z,S=k$ .

The distributions in \eqref{eq:prior1}-\eqref{eq:prior3} are conjugate priors that allow us to obtain the posterior distributions for the parameters involved conveniently. Consequently, we can estimate them with an efficient Gibbs sampling algorithm. 

The Gibbs sampling algorithm was outlined in Algorithm \ref{alg:model}. In each iteration $r=1,\dots,R$, it first draws from the posterior distributions of the parameters and then computes $E[Y(x)]$ following the result in \eqref{eqn: y(x)}.
Below are the detailed steps:

\begin{itemize}
    \item Draw from the posterior distributions  for the parameters in the model of $Y|X,Z, S$ and the latent variables $S$ involved:
    \begin{itemize}
        \item \textit{Component allocation:}  the posterior distribution of the latent variables $S_{i}$ follows a multinomial distribution: 
    \begin{equation}
        Pr(S_{i}=k) \propto \omega_{k}(x_i)\mathcal{N}(y_i;\theta_{0,k}+\theta_{X,k}x_i+\theta_{Z,k}z_i,\delta_{y,k}^{2}), \mbox{ for } k=1,\dots,K;\notag 
    \end{equation}
with $\omega_{k}(x_i) = \Phi(\alpha_k(x_i)) \prod_{r<k} (1-\Phi(\alpha_r(x_i))),$ for $ k=1,\dots,K$ and $i=1,\dots,n$, where    $\Phi(\alpha_k(x_i))=1$ for k =K.

        \item \textit{Component-specific parameters:}
        Conditional on the latent variables $\{S_{i}\}_{i=1}^n$, we can update the values of the parameters from their posterior distributions, for $k=1,\dots,K$:
        \begin{align*}
        {\boldsymbol\theta}_{Y,k} & \sim \mathcal{N}\left(V_k\cdot \left( \frac{\mu_{\theta}}{\sigma_{\theta}^2}+\frac{\sum_{i \in \{ S_{i}=k\}} y_i\cdot[1,x_{i: S_{i}=k},z_{i: S_{i}=k}]^T)}{\delta^2_{y,k}} \right),V_k \right)\\
     \mbox{ where } V_k &= \frac{1}{\sigma_{\theta}^2}\cdot \mathbf{1}_3 +\frac{[1,x_{i: S_{i}=k},z_{i: S_{i}=k}]^T[1,x_{i: S_{i}=k},z_{i: S_{i}=k}]}{\delta_{y,k}^{2}},\\
        \delta_{y,k}^{2} & \sim \mbox{InvGamma}\left(\gamma_1+ \frac{n_{k}}{2},\gamma_2+\frac{\sum_{i : S_{i}=k} [y_{i}-(\theta_{0,k}+\theta_{X,k}x_i+\theta_{Z,k}z_i)]^2}{2}\right),
        \end{align*}
        and $n_k$ is the number of units allocated to the mixture component $k$.

        \item \textit{Augmentation scheme:} In order to sample from $\{\alpha_k(x)\}_{k=1}^K$ we use the data augmentation scheme, developed by Albert and Chib\cite{albert2001sequential} and borrowed by Rodriguez and Dunson \cite{rodriguez2011nonparametric}. We can impute the augmented variables $\{ Q_k(x_i)\}_{k=1}^{\min(S_{i},K-1)}$, for each $i$, by sampling from its full conditional distribution:
        \[
        Q_{k}(x_i)|S_{i},\alpha_k(x_i) \sim \begin{cases}
        \mathcal{N}(\alpha_k(x_i),1)\mathbf{I}_{\mathrm{R}^+} \mbox{ if } k=S_{i},\\
        \mathcal{N}(\alpha_k(x_i),1)\mathbf{I}_{\mathrm{R}^-} \mbox{ if } k<S_{i};\end{cases}
        \]
        where $\alpha_k(x_i)$ is updated using the the new values of $\omega_k(x_i)$ by
        \[
        \alpha_k(x_i)=\phi\left(\frac{\omega_{k}(x_i)}{\prod_{r<k}(1-\Phi(x_i^T{\boldsymbol\eta}_{r})}\right)= \phi\left(\frac{\omega_{k}(x_i)}{1-\sum_{r<k}\omega_{r}(x_i)}\right);
        \]
        $\phi(\cdot)$ is the density function of the Gaussian distribution. 

        \item \textit{Weights:} The posterior distribution for ${\boldsymbol\eta}_{k}$, for  $k=1,\dots,\max(S_{i},K-1)$, is:
        \[
        {\boldsymbol\eta}_{k} \sim \mathcal{N}\left( \left(\frac{1}{\sigma^2_{\eta}}\mathbf{1}_p+
    \Tilde{x}_k^T\Tilde{x}_k\right)^{-1}\cdot \left(\frac{\mu_\eta}{\sigma_\eta^2} +\Tilde{x}_k^T\Tilde{Q}_k\right),\left(\frac{1}{\sigma^2_{\eta}}\mathbf{1}_p+\Tilde{x}_k^T\Tilde{x}_k\right)^{-1}\right) 
        \]
        where $\Tilde{x}_k$ is a matrix  composed by the rows  $[1,x_i]$,  $i \in \{ S_{i}\leq k \}$ in model \eqref{weight_1} and $\Tilde{Q}_k$ is a vector composed by the $Q_{k}(x_i)$, $i \in \{ S_{i}\leq k \}$.
    \end{itemize}

    \item Draw from the posterior distribution of the parameters in the $W|X,Z$ model:
    \begin{align*}
        \boldsymbol{\theta}_W &\sim \mathcal{N}_3 \biggl(V^{-1}\biggl(\frac{1}{\sigma^2_\theta} \mu_\theta + \frac{1}{\delta^2_w}[1,x,z]^T w \biggl),V^{-1} \biggl)\\
     V &=\frac{1}{\sigma^2_\theta} \mathbf{1}_3 + \frac{1}{\gamma^2_w} [1,x,z]^T [1,x,z];\\
    \delta^2_w & \sim \mbox{InvGamma} \left(\gamma_3 +\frac{n}{2},\gamma_3 +\frac{\sum_{i=1}^n (w_i - (\theta_{W0}  + \theta_{WX} x_i +\theta_{WZ} z_i))^2}{2}\right).
    \end{align*}
    $n$ is the total sample size and $x$, $z$, $w$ are the observed values of $n$-dimension for variables $X$, $Z$, and $W$.

    \item According to \eqref{eqn: beta_k}-\eqref{eqn: intercept}, identify  the slope parameter $\beta_{x,k}$ and the intercept $\iota_{k}$  for $k=1,\dots,K$ as follows:
    \begin{align*}
        \beta_{x,k} &= \theta_{X,k}-\theta_{Z,k}\frac{\theta_{WX}}{\theta_{WZ}}, \\
        \iota_{k} &= \theta_{0,k} +\theta_{Z,k}\cdot\sum_{i=1}^n \frac{z_i}{n} + \theta_{Z,k}\frac{\theta_{WX}}{\theta_{WZ}}\cdot\sum_{i=1}^n \frac{x_i}{n}. 
    \end{align*}
\end{itemize}

%% file: Gibbs.tex
\begin{algorithm}
\caption{Estimation Model}
\label{alg:model}
\vspace{0.15cm}
{\bf Inputs:} the observed data $(y,x,z,w)$.

{\bf Outputs:} posterior distributions of parameters.

{\bf Procedure:}
\begin{algorithmic}
\State Choice of hyperparameters;
\State Initialization of:
\State - parameters for Y-model: ${\boldsymbol\theta_{Y,k}}$, $\delta_{y,k}$, ${\boldsymbol\eta_k}$, for $k=1,\dots,K$;
\State - parameters for W-model: ${\boldsymbol\theta_W}$,  $\delta_w$;
\State - latent variables: $S_{i}$ and $Q_k(x_i)$ for $i=1,\dots,n$ and $k=1,\dots, K-1$ \textit{(see Appendix A)}.
\For{$r \in \{1,\dots,R\}$}
    \State $\xrightarrow{}$ \textit{Parameters for $Y|X,Z$ model:}
     \State Draw $S_{i}$ for $i=1,\dots,n$;
    \State Draw ${\boldsymbol\theta_{Y,k}}$ and  $\delta_{y,k}$, for $k=1,\dots,K$;
     \State Draw ${\boldsymbol\eta_k}$ for $k=1,\dots,K$;
     \State Compute $\alpha_k(x_i)$ for $i=1,\dots,n$;
    \State Draw $Q(x_i)$ for $i=1,\dots,n$; \textit{(see Appendix A)}
     \State Compute $\omega_k(x_i)$ for $i=1,\dots,n$ and $k=1,\dots,K$;
    \State $\xrightarrow{}$ \textit{Parameters for $W|X,Z$ model:}
    \State Draw ${\boldsymbol\theta_W}$ and  $\delta_w$;
    \State $\xrightarrow{}$ \textit{Parameters for the CERF:}
    \State Compute $E[Y(x)]$.
\EndFor
\end{algorithmic}
\end{algorithm}

%% file: Additional_simulations.tex
\begin{figure}[hbp!]
\begin{center}
\includegraphics[width=2.2in]{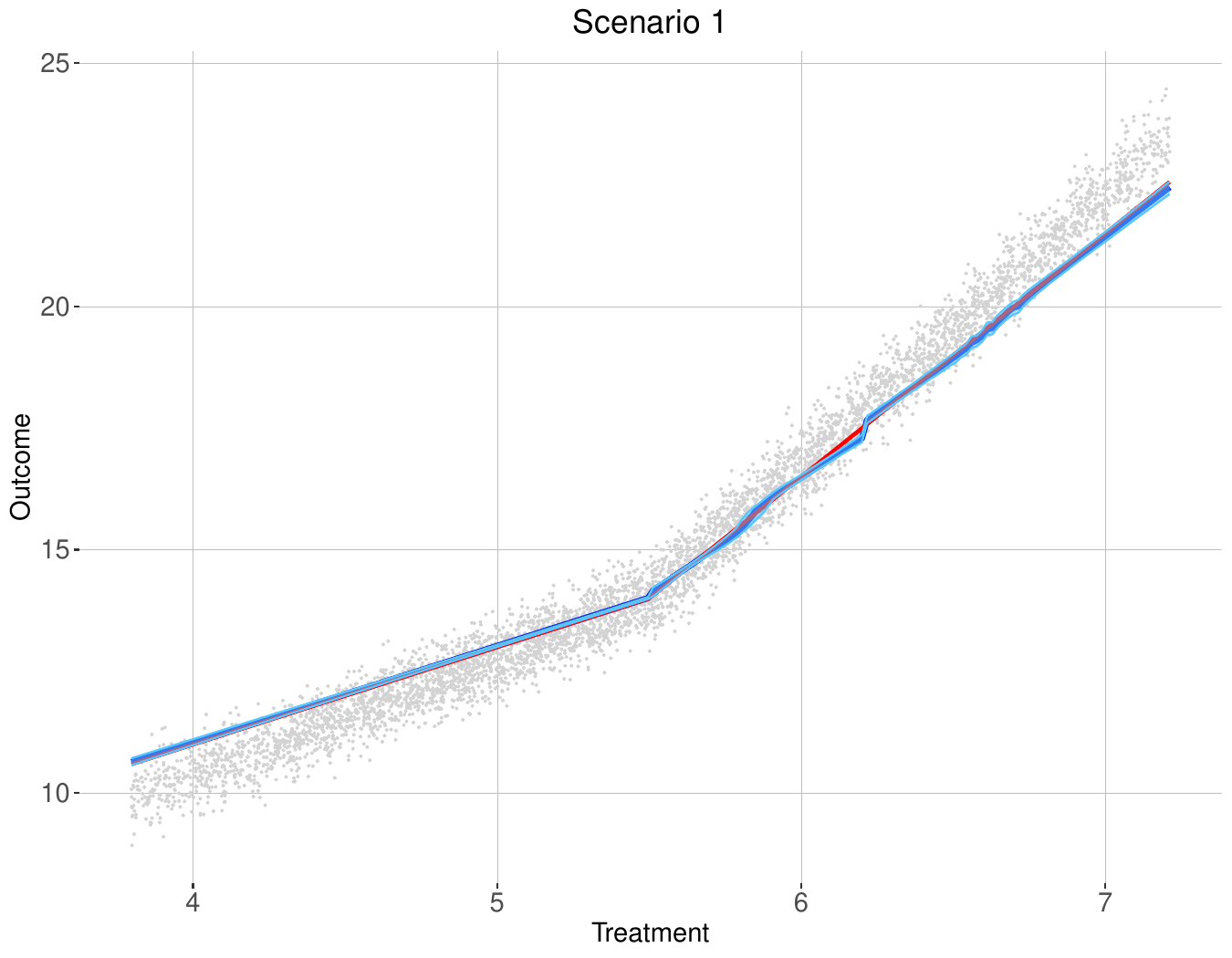}\,\includegraphics[width=2.2in]{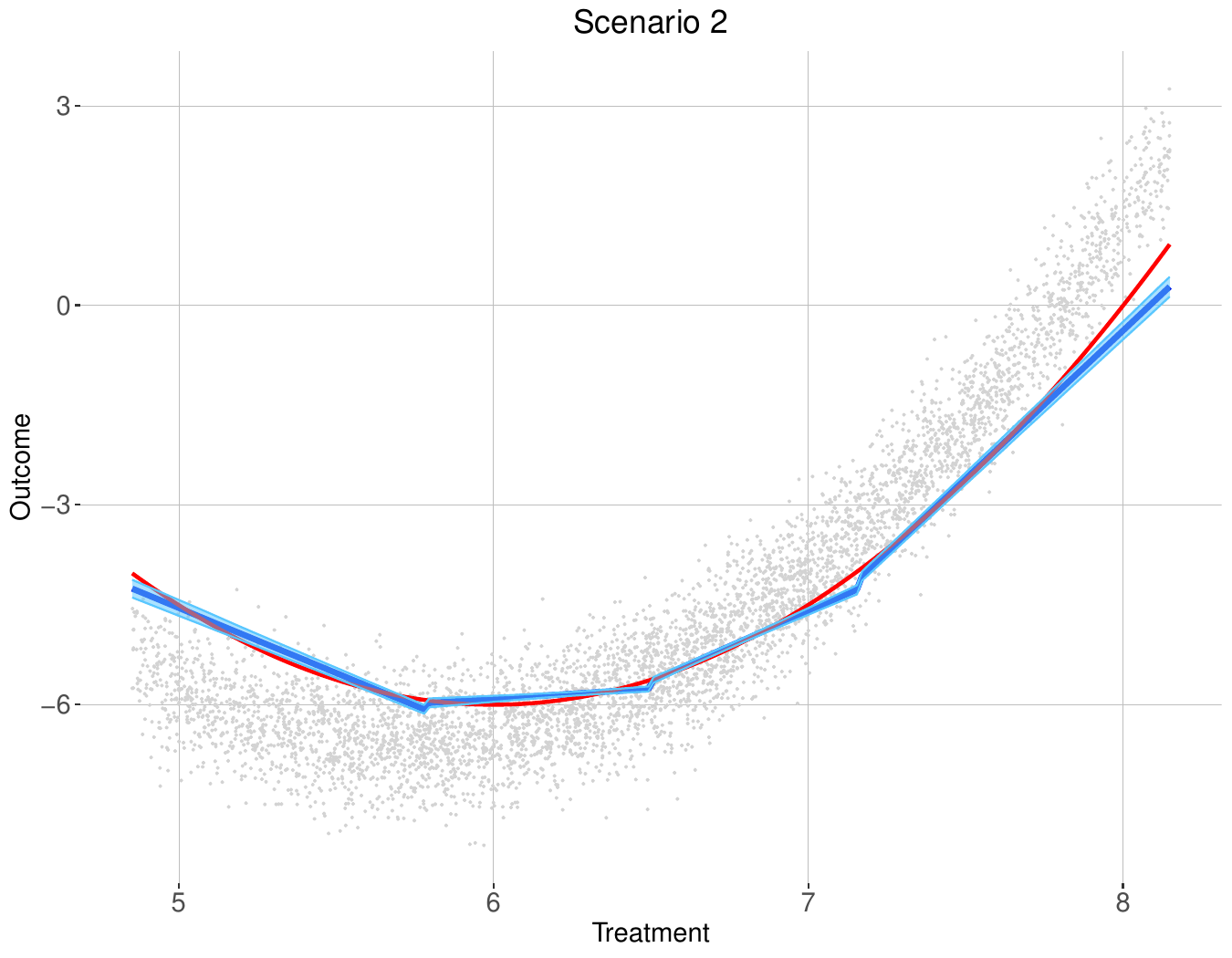}\\
\includegraphics[width=2.2in]{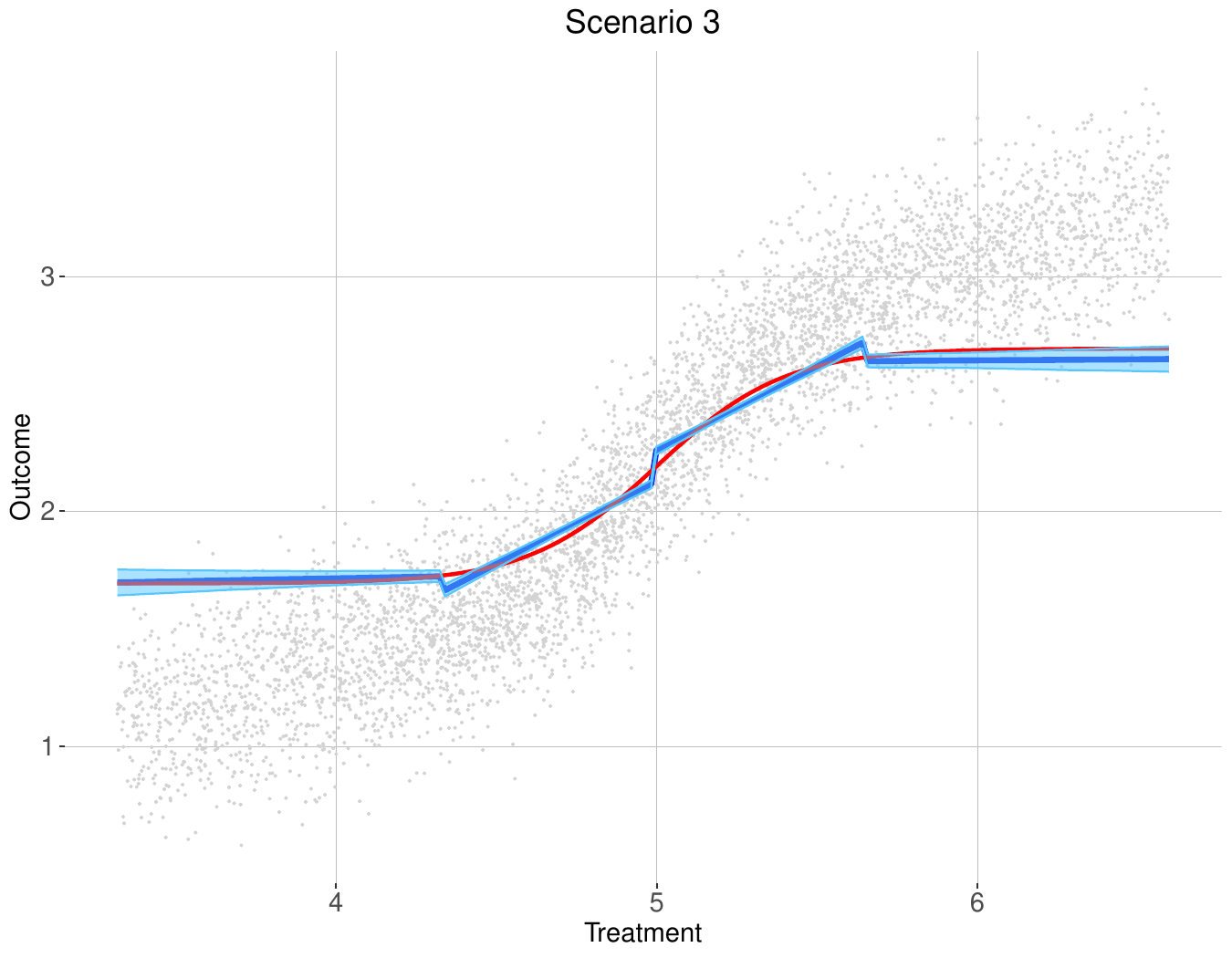}\,\includegraphics[width=2.2in]{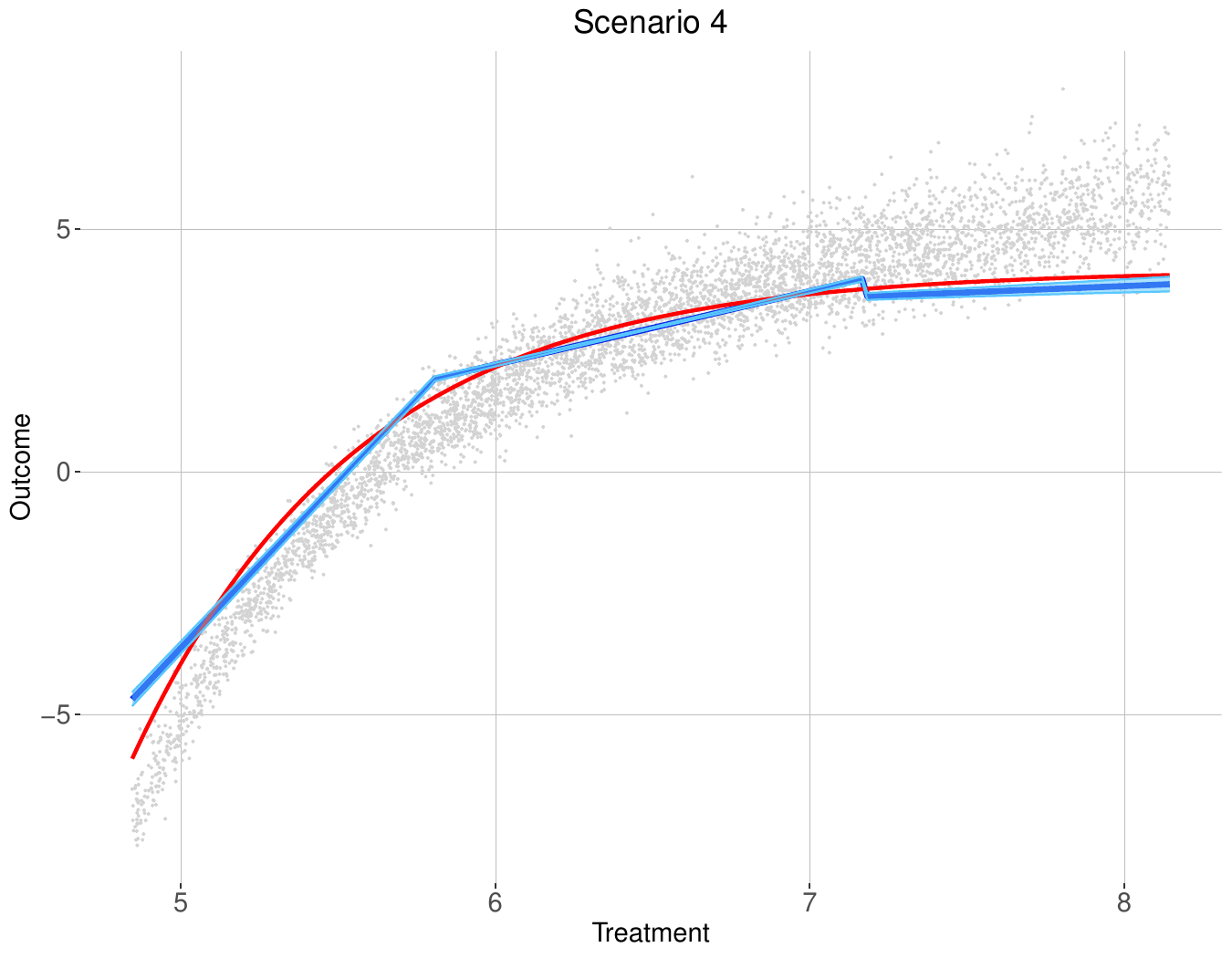}
\end{center}
\caption{Comparison of the median of the estimated CERFs posterior without kernel smoothing (the blue line) to the true CERF(the red line) under four scenarios and data setting A using weight prior model \eqref{weight_2} is.  Light blue bands represent the width of 95\% credible intervals. Light grey dots show the simulated data.}
\label{fig:no_smoothing}
\end{figure}

\begin{figure}[tbp]
\begin{center}
\includegraphics[width=2.2in]{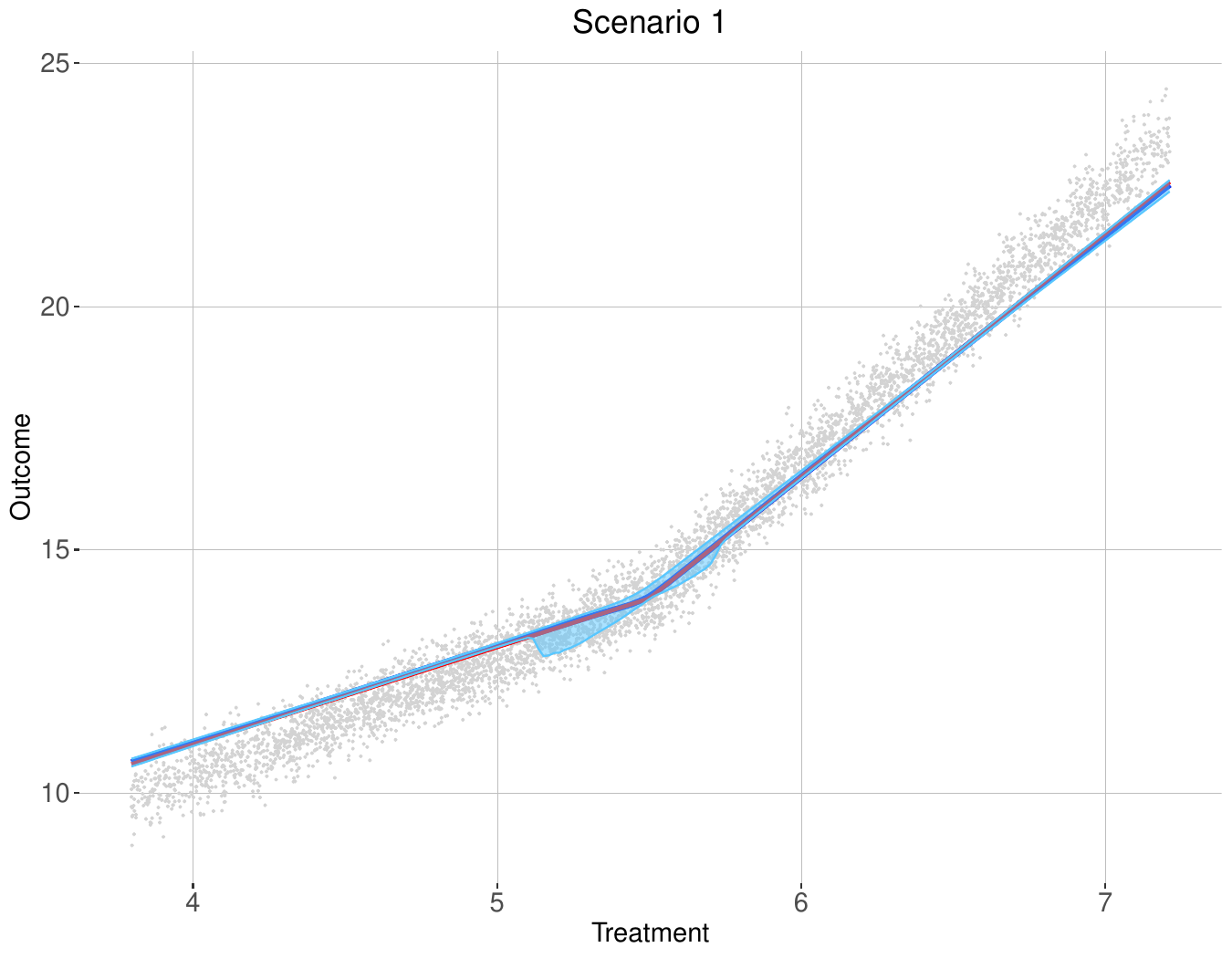}\,\includegraphics[width=2.2in]{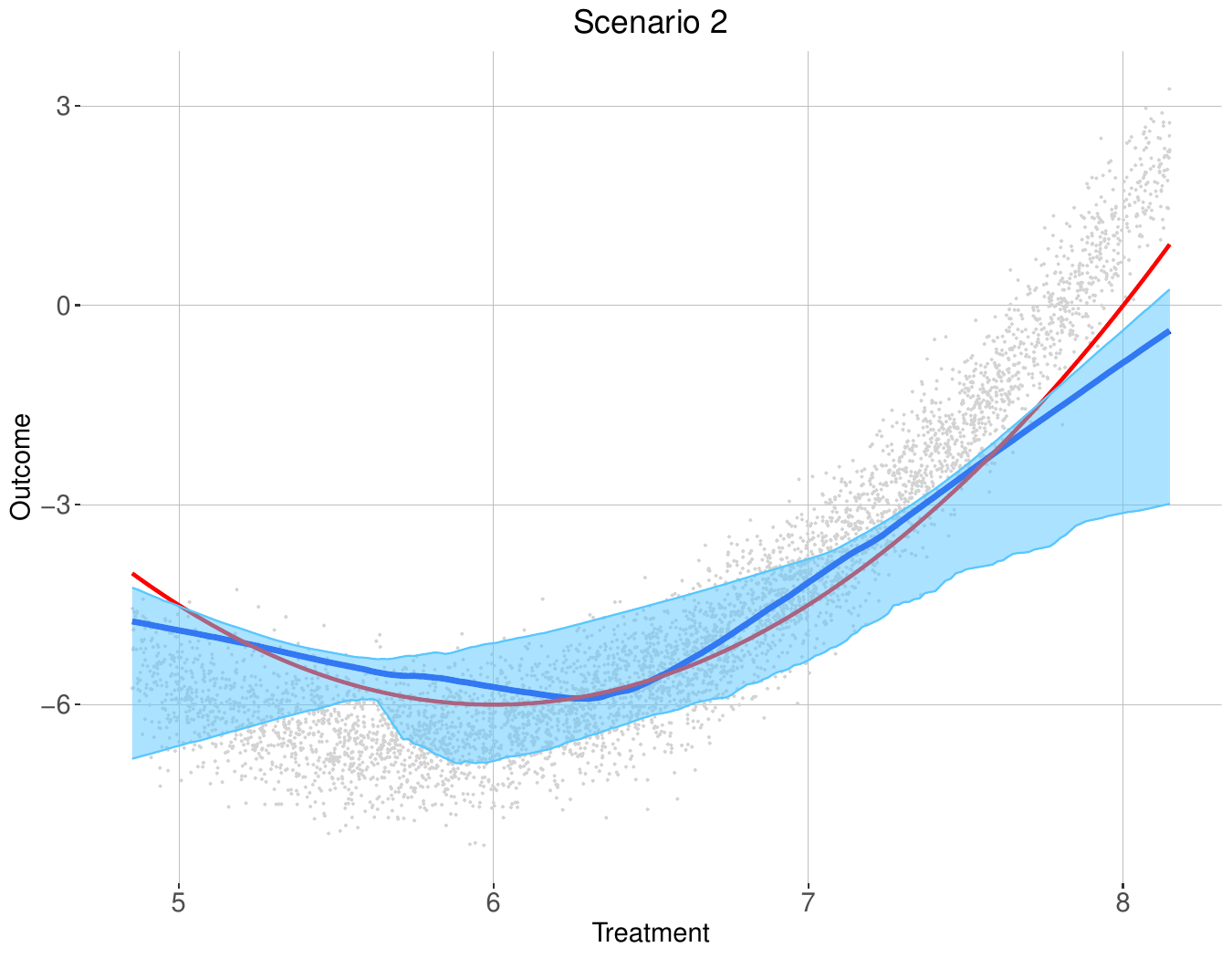}\\
\includegraphics[width=2.2in]{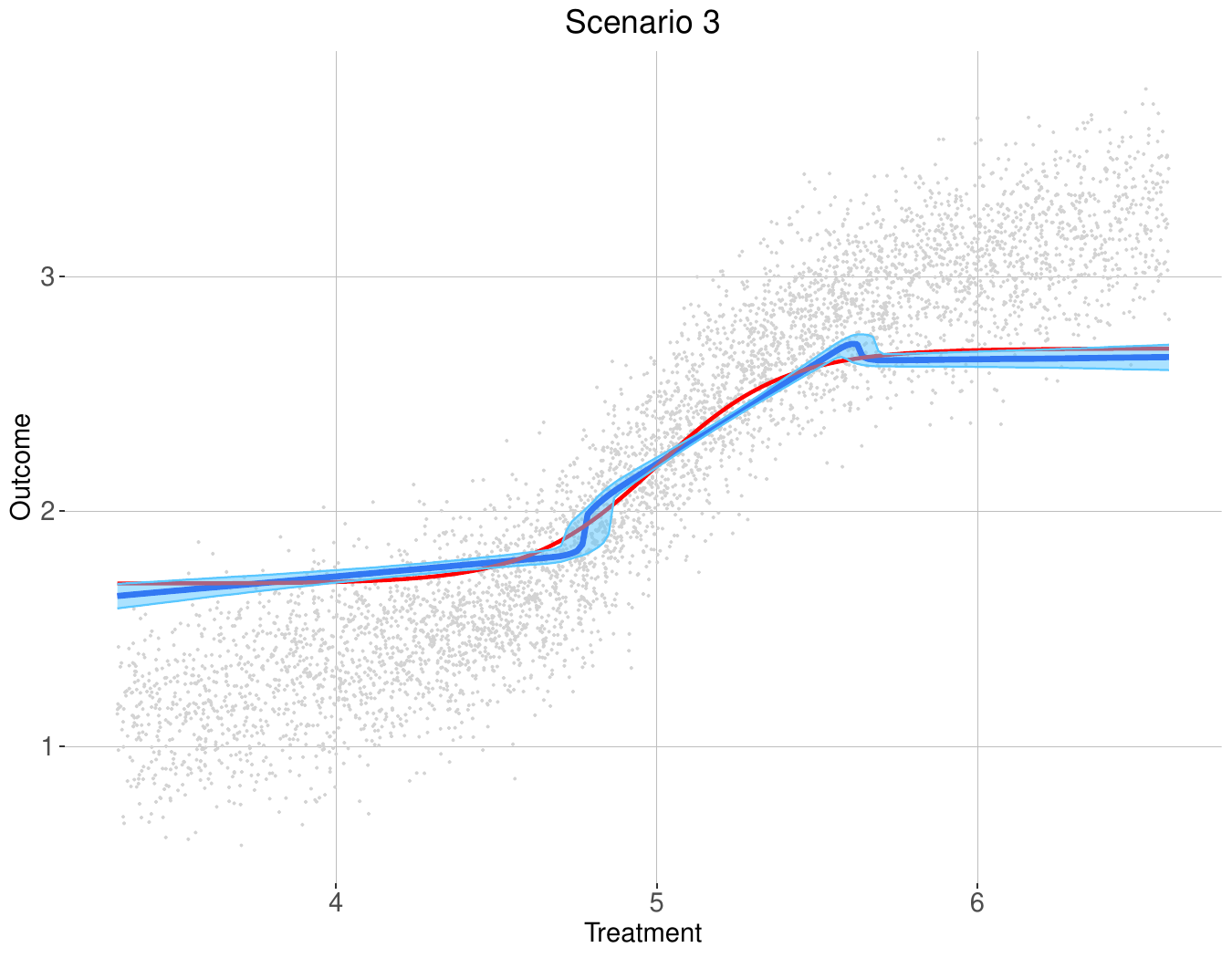}\,\includegraphics[width=2.2in]{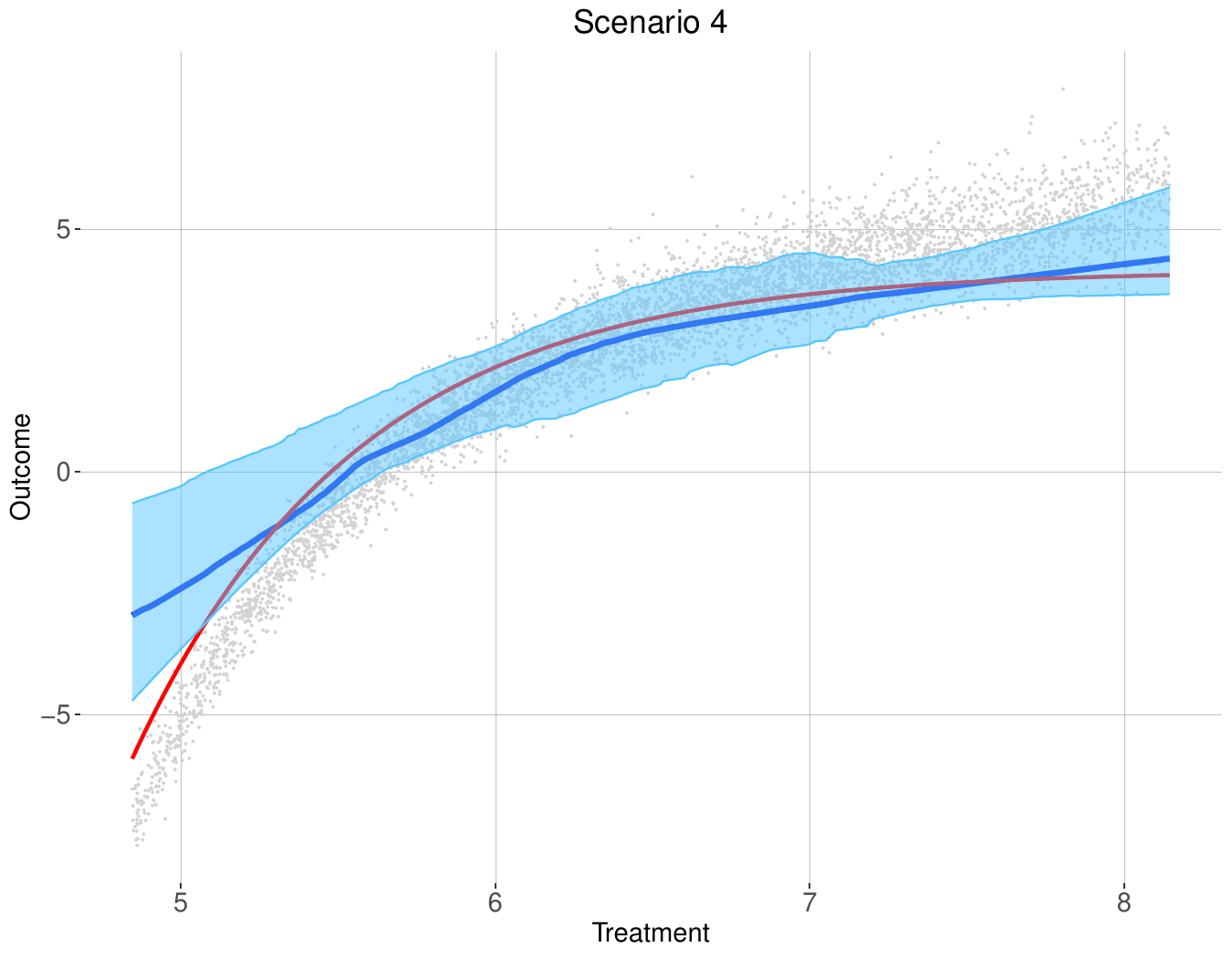}
\end{center}
\caption{Comparison of the median of the estimated CERFs posterior (the dark blue line) to the true CERF(the red line) under four scenarios of varying CERF shapes using weight prior model \eqref{weight_1}.  Light blue bands represent the width of 95\% credible intervals. Light grey dots show the simulated data. Sample size is $5000$.}
\label{fig:no_splitting}
\end{figure}